%

%
\input harvmac.tex
\input amssym.def

\def\CD{{\cal D}}

\def\CN{{\cal N}}
\def\CO{{\cal O}}

\def\V{{\frak V}}

\def\hf{{\hat f}}
\def\hg{{\hat g}}

\def\Db{{\bar D}}
\def\Ab{{\bar A}}
\def\Bb{{\bar B}}
\def\Cb{{\bar C}}

\def\hnab{{\hat \nabla}}

\def\hbnab{{\hat{\bar\nabla}}}
\def\bnabla{{\bar \nabla}}
\def\bfnabla{{\nabla\kern-.87em\nabla}}

\def\ebar{{\bar \epsilon}}
\font\cmss=cmss10 \font\cmsss=cmss10 at 7pt
\def\IR{\relax{\rm I\kern-.18em R}}
\def\inbar{\vrule height1.5ex width.4pt depth0pt}
\def\IC{\relax\,\hbox{$\inbar\kern-.3em{\rm C}$}}
\def\IN{\relax{\rm I\kern-.18em N}}
\def\IF{\relax{\rm I\kern-.18em F}}
\def\IP{\relax{\rm I\kern-.18em P}}
\def\IZ{\relax\ifmmode\mathchoice
{\hbox{\cmss Z\kern-.4em Z}}{\hbox{\cmss Z\kern-.4em Z}}
{\lower.9pt\hbox{\cmsss Z\kern-.4em Z}} {\lower1.2pt\hbox{\cmsss
Z\kern-.4em Z}}\else{\cmss Z\kern-.4em Z}\fi}
\def\noth#1{\hbox{$#1$\kern-1.0em\raise2.8ex\hbox{$_\circ$}}}
\def\nothi#1{\hbox{$#1$\kern-.4em\raise2.7ex\hbox{$_\circ$}}}
\def\nothii#1{\hbox{$#1$\kern-.4em\raise2.0ex\hbox{$_\circ$}}}
\def\nothiii#1{\hbox{$#1$\kern-.4em\raise2.5ex\hbox{$_\circ$}}}
\def\nc{\hbox{$\nabla$\kern-.57em\raise2.4ex\hbox{$_\circ$}}}
\def\nhc{\hbox{${\hat\nabla}$\kern-.57em\raise2.7ex\hbox{$_\circ$}}}
\def\Wc{\hbox{$W$\kern-.65em\raise2.4ex\hbox{$_\circ$}}}
\def\Whc{\hbox{${\hat W}$\kern-.65em\raise2.8ex\hbox{$_\circ$}}}
\def\cc{\hbox{$\chi$\kern-.5em\raise1.8ex\hbox{$_\circ$}}}
\def\chc{\hbox{${\hat\chi}$\kern-.5em\raise2.2ex\hbox{$_\circ$}}}
\def\bcc{\hbox{${\bar \chi}$\kern-.5em\raise2ex\hbox{$_\circ$}}}
\def\bchc{\hbox{${\hat{\bar\chi}}$\kern-.5em\raise2.5ex\hbox{$_\circ$}}}
\def\sm#1#2{\kern.1em\lower.4ex\hbox{${\scriptstyle #2}$\kern-1em\raise1.6ex
             \hbox{${\scriptstyle #1}$}}}

\def\b{{\beta}}
\def\bd{{\dot{\beta}}}
\def\a{{\alpha}}
\def\ad{{\dot{\alpha}}}
\def\d{{\delta}}
\def\dd{{{\dot\delta}}}
\def\g{{\gamma}}
\def\gd{{\dot{\gamma}}}
\def\o{{\omega}}
\def\O{{\Omega}}
\def\bo{{\bar \omega}}
\def\bchi{{\bar\chi}}
\def\s{{\sigma}}
\def\t{{\theta}}
\def\bt{{\bar\theta}}
\def\l{{\lambda}}
\def\L{{\Lambda}}
\def\hL{{\hat\Lambda}}
\def\sfrac#1#2{\textstyle{#1\over #2}}
\def\contra#1#2{\,{\buildrel \,\hbox{$
    \vrule height 3pt width 1pt depth 0pt
    \vrule height 3pt width #1pt depth -2pt
    \vrule height 3pt width 1pt depth 0pt $}
    \over {#2} }\,}

%
%

\lref\BrittoAJ{R.~Britto, B.~Feng and S.~J.~Rey,
      {\it Deformed superspace, $\CN =1/2$ supersymmetry and 
           (non)renormalization theorems},
      JHEP {\bf 0307}, 067 (2003)
      [hep-th/0306215].
}

\lref\BrittoAJj{R.~Britto, B.~Feng and S.~J.~Rey,
      {\it Non(anti)commutative superspace, UV/IR mixing and open Wilson 
           lines},
      JHEP {\bf 0308}, 001 (2003)
      [hep-th/0307091].
}

\lref\BerensteinSR{D.~Berenstein and S.~J.~Rey,
      {\it Wilsonian proof for renormalizability of $\CN=1/2$ supersymmetric 
           field theories},
      Phys.\ Rev.\ D {\bf 68}, 121701 (2003)
      [hep-th/0308049].
}

\lref\WessCP{J.~Wess and J.~Bagger,
      {\it Supersymmetry and supergravity},
      Princeton University Press, Princeton (1992).
}

\lref\FerraraXY{S.~Ferrara, M.~A.~Lledo and O.~Macia,
      {\it Supersymmetry in noncommutative superspaces},
      JHEP {\bf 0309}, 068 (2003)
      [hep-th/0307039].
}

\lref\HarnadBC{J.~P.~Harnad and S.~Shnider,
      {\it Constraints and field equations for 
            ten-dimensional super Yang-Mills theory},
      Commun.\ Math.\ Phys.\  {\bf 106}, 183 (1986).
}

\lref\HarnadVK{J.~P.~Harnad, J.~Hurtubise, M.~Legare and S.~Shnider,
      {\it Constraint equations and field equations in 
           supersymmetric $\CN=3$ Yang-Mills theory},
      Nucl.\ Phys.\ B {\bf 256}, 609 (1985).
}

\lref\VanProeyenNI{A.~Van Proeyen,
      {\it Tools for supersymmetry},
      hep-th/9910030.
}

\lref\SeibergVS{N.~Seiberg and E.~Witten,
      {\it String theory and noncommutative geometry},
      JHEP {\bf 9909}, 032 (1999)
      [hep-th/9908142].
}

\lref\BerkovitsKJ{
      N.~Berkovits and N.~Seiberg,
      {\it Superstrings in graviphoton background and 
        $\CN = 1/2 + 3/2$ supersymmetry},
      JHEP {\bf 0307}, 010 (2003)
      [hep-th/0306226].
}

\lref\JurcoRQ{
      B.~Jurco, L.~M\"oller, S.~Schraml, P.~Schupp and J.~Wess,
      {\it Construction of non-Abelian gauge theories on 
            noncommutative spaces},
      Eur.\ Phys.\ J.\ C {\bf 21}, 383 (2001)
      [hep-th/0104153].
}

\lref\SeibergYZ{N.~Seiberg,
      {\it Noncommutative superspace, $\CN = 1/2$ 
            supersymmetry, field theory and  string theory},
      JHEP {\bf 0306}, 010 (2003)
      [hep-th/0305248].
}

\lref\deBoerDN{
      J.~de Boer, P.~A.~Grassi and P.~van Nieuwenhuizen,
      {\it Noncommutative superspace from string theory},
      Phys.\ Lett.\ B {\bf 574}, 98 (2003)
      [hep-th/0302078].
}

\lref\OoguriQP{
      H.~Ooguri and C.~Vafa,
      {\it The $C$-deformation of gluino and non-planar 
       diagrams},
       Adv.\ Theor.\ Math.\ Phys.\  {\bf 7}, 53 (2003)
       [hep-th/0302109].
}

\lref\OoguriTT{
      H.~Ooguri and C.~Vafa,
      {\it Gravity induced C-deformation},
      hep-th/0303063.
}

\lref\IvanovTE{
      E.~Ivanov, O.~Lechtenfeld and B.~Zupnik,
      {\it Nilpotent deformations of $\CN=2$ superspace},
      hep-th/0308012.
}

\lref\BrinkNB{
      L.~Brink and J.~H.~Schwarz,
      {\it Quantum superspace},
      Phys.\ Lett.\ B {\bf 100}, 310 (1981).
}

\lref\FerraraMM{
      S.~Ferrara and M.~A.~Lledo,
      {\it Some aspects of deformations of supersymmetric 
      field theories},
      JHEP {\bf 0005}, 008 (2000)
      [hep-th/0002084].
}

\lref\KosinskiXU{
      P.~Kosinski, J.~Lukierski and P.~Maslanka,
      {\it Quantum deformations of space-time SUSY and 
      noncommutative superfield theory},
      hep-th/0011053.
}

\lref\KlemmYU{
      D.~Klemm, S.~Penati and L.~Tamassia,
      {\it Non(anti)commutative superspace},
      Class.\ Quant.\ Grav.\  {\bf 20}, 2905 (2003)
      [hep-th/0104190].
}

\lref\BuchbinderAT{
      I.~L.~Buchbinder and I.~B.~Samsonov,
      {\it Noncommutative $\CN = 2$ supersymmetric theories in 
      harmonic superspace},
      Grav.\ Cosmol.\  {\bf 8}, 17 (2002)
      [hep-th/0109130].
}

\lref\ImaanpurJJ{
      A.~Imaanpur,
      {\it On instantons and zero modes of $\CN = 1/2$ SYM 
       theory},
      JHEP {\bf 0309}, 077 (2003)
      [hep-th/0308171].
}

\lref\GrassiQK{
      P.~A.~Grassi, R.~Ricci and D.~Robles-Llana,
      {\it Instanton calculations for $\CN = 1/2$
        super Yang-Mills theory},
      hep-th/0311155.
}

\lref\BrittoUV{
      R.~Britto, B.~Feng, O.~Lunin and S.~J.~Rey,
      {\it $U(N)$ instantons on $\CN = 1/2$ superspace: 
       Exact solution and geometry of moduli space},
      hep-th/0311275.
}

\lref\ArakiMQ{
      T.~Araki, K.~Ito and A.~Ohtsuka,
      {\it $\CN=2$ supersymmetric $U(1)$ gauge theory in 
      noncommutative harmonic superspace},
      hep-th/0401012.
}

\lref\GrisaruFD{
     M.~T.~Grisaru, S.~Penati and A.~Romagnoni,
     {\it Two-loop renormalization for nonanticommutative 
      $\CN = 1/2$ supersymmetric WZ model},
     JHEP {\bf 0308}, 003 (2003)
     [hep-th/0307099].
}

\lref\BrittoKG{
      R.~Britto and B.~Feng,
      {\it $\CN = 1/2$ Wess-Zumino model is renormalizable},
      Phys.\ Rev.\ Lett.\  {\bf 91}, 201601 (2003)
      [hep-th/0307165].
}

\lref\LuninBM{
      O.~Lunin and S.~J.~Rey,
      {\it Renormalizability of non(anti)commutative gauge 
       theories with $\CN = 1/2$ supersymmetry},
      JHEP {\bf 0309}, 045 (2003)
      [hep-th/0307275].
}

\lref\AlishahihaKG{
      M.~Alishahiha, A.~Ghodsi and N.~Sadooghi,
      {\it One-loop perturbative corrections to 
      non(anti)commutativity parameter of $\CN =1/2$ supersymmetric $U(N)$
       gauge theory},
      hep-th/0309037.
}

\lref\WittenNT{
      E.~Witten,
      {\it Twistorlike transform in ten dimensions},
      Nucl.\ Phys.\ B {\bf 266}, 245 (1986).
}

\lref\Zakharov{
      V.~E.~Zakharov and A.~B.~Shabat,
      {\it Integration of non-linear equations of mathematical physics by the
      method of inverse scattering},
      Funct.\ Anal.\ Appl.\ {\bf 13}, 166 (1979).
}

\lref\WardTA{
      R.~S.~Ward,
      {\it On self-dual gauge fields},
      Phys.\ Lett.\ A {\bf 61}, 81 (1977).
}

\lref\FerraraXK{
      S.~Ferrara and E.~Sokatchev,
      {\it Nonanticommutative $\CN = 2$ super Yang-Mills theory with singlet
      deformation},
      Phys.\ Lett.\ B {\bf 579}, 226 (2004)
      [hep-th/0308021].
}

\lref\VolovichKR{
      I.~V.~Volovich,
      {\it Supersymmetric Yang-Mills equations as an inverse scattering 
      problem},
      Lett.\ Math.\ Phys.\  {\bf 7}, 517 (1983).
}

\lref\AbdallaXQ{
      E.~Abdalla, M.~Forger and M.~Jacques,
      {\it Higher conservation laws for ten-dimensional supersymmetric 
      Yang-Mills theories},
      Nucl.\ Phys.\ B {\bf 307}, 198 (1988).
}

\lref\ChauIV{
      L.~L.~Chau and B.~Milewski,
      {\it Linear system and conservation laws for $D=10$ super Yang-Mills},
      Phys.\ Lett.\ B {\bf 198}, 356 (1987).
}

\lref\Kapranov{
      M.~M.~Kapranov and Yu.~I.~Manin,
      {\it Twistor transform and algebraic geometry constructions of solutions
      to field theory equations},
      Russian\ Math.\ Surveys {\bf 41:5}, 85 (1986).
}

\lref\GervaisHH{
      J.~L.~Gervais and M.~Savelev,
      {\it Progress in classically solving ten-dimensional supersymmetric 
      reduced Yang-Mills theories},
      Nucl.\ Phys.\ B {\bf 554}, 183 (1999)
      [hep-th/9811108].
}

\lref\GervaisVJ{
      J.~L.~Gervais and H.~Samtleben,
      {\it Integrable structures in classical off-shell $10D$ supersymmetric  
      Yang-Mills theory},
      Commun.\ Math.\ Phys.\  {\bf 217}, 1 (2001)
      [hep-th/9912089].
}

\lref\LechtenfeldAW{
      O.~Lechtenfeld and A.~D.~Popov,
      {\it Noncommutative multi-solitons in $2+1$ dimensions},
      JHEP {\bf 0111}, 040 (2001)
      [hep-th/0106213].
}

\lref\LechtenfeldGF{
      O.~Lechtenfeld and A.~D.~Popov,
      {\it Scattering of noncommutative solitons in $2+1$ dimensions},
      Phys.\ Lett.\ B {\bf 523}, 178 (2001)
      [hep-th/0108118].
}

\lref\LechtenfeldIE{
      O.~Lechtenfeld and A.~D.~Popov,
      {\it Noncommutative 't Hooft instantons},
      JHEP {\bf 0203}, 040 (2002)
      [hep-th/0109209].
}

\lref\WolfJW{
      M.~Wolf,
      {\it Soliton-antisoliton scattering configurations in a noncommutative 
      sigma model in $2+1$ dimensions},
      JHEP {\bf 0206}, 055 (2002)
      [hep-th/0204185].
}

\lref\HorvathBJ{
      Z.~Horvath, O.~Lechtenfeld and M.~Wolf,
      {\it Noncommutative instantons via dressing and splitting approaches},
      JHEP {\bf 0212}, 060 (2002)
      [hep-th/0211041].
}

\lref\IhlKZ{
      M.~Ihl and S.~Uhlmann,
      {\it Noncommutative extended waves and soliton-like configurations in 
      $\CN = 2$ string theory},
      Int.\ J.\ Mod.\ Phys.\ A {\bf 18}, 4889 (2003)
      [hep-th/0211263].
}

\lref\LechtenfeldVV{
      O.~Lechtenfeld and A.~D.~Popov,
      {\it Noncommutative monopoles and Riemann-Hilbert problems},
      hep-th/0306263.
}

\lref\GaillardMN{
      M.~K.~Gaillard,
      {\it Pauli-Villars regularization of globally supersymmetric theories},
      Phys.\ Lett.\ B {\bf 347}, 284 (1995)
      [hep-th/9412125].
}

\lref\BerkovitsUE{
      N.~Berkovits and P.~S.~Howe,
      {\it Ten-dimensional supergravity constraints from the pure spinor 
      formalism  for the superstring},
      Nucl.\ Phys.\ B {\bf 635}, 75 (2002)
      [hep-th/0112160].
}

\lref\DayiJU{O.~F.~Dayi, K.~Ulker and B.~Yapiskan,
      {\it Duals of noncommutative supersymmetric $U(1)$ gauge theory},
      JHEP {\bf 0310}, 010 (2003)
      [hep-th/0309073].
}

\lref\MikulovicSQ{D.~Mikulovic,
      {\it Seiberg-Witten map for superfields on canonically deformed 
           $\CN = 1$, $d = 4$ superspace},
      hep-th/0310065.
}

\lref\LukierskiJW{
      J.~Lukierski and W.~J.~Zakrzewski,
      {\it Euclidean supersymmetrization of instantons and self-dual 
      monopoles},
      Phys.\ Lett.\ B {\bf 189}, 99 (1987).
}

\lref\Saemann{
      O.~Lechtenfeld, C.~S\"amann and M.~Wolf,
      {\it work in progress.}
}

\lref\ArakiSE{
      T.~Araki, K.~Ito and A.~Ohtsuka,
      {\it Supersymmetric gauge theories on noncommutative superspace,}
      Phys.\ Lett.\ B {\bf 573}, 209 (2003)
      [hep-th/0307076].
}

\newbox\tmpbox\setbox\tmpbox\hbox{\abstractfont }

\Title{ \vbox{
\rightline{\hbox{hep-th/0401147}}
\rightline{\hbox{ITP--UH--05/04}}}}
{\vbox{\centerline{Constraint and Super Yang-Mills Equations}
\bigskip
\centerline{on the Deformed Superspace $\IR^{(4|16)}_\hbar$}}}
\smallskip
\centerline{Christian S\"amann and Martin Wolf
\foot{\tt saemann, wolf@itp.uni-hannover.de}
}

\smallskip
\centerline{\it Institut f\"ur Theoretische Physik}
\centerline{\it Universit\"at Hannover}
\centerline{\it Appelstra{\ss}e 2, 30167 Hannover, Germany}

\bigskip
\vskip 1cm
\centerline{\bf Abstract} 
It has been known for quite some time that the $\CN=4$ super 
Yang-Mills equations defined on four-dimensional Euclidean space are 
equivalent to certain constraint equations on the  
Euclidean superspace $\IR^{(4|16)}$. In this paper we 
consider the constraint equations on a deformed superspace 
$\IR^{(4|16)}_\hbar$ \`a la Seiberg and derive the
deformed super Yang-Mills equations. In showing this, 
we propose a super Seiberg-Witten map.

\Date{January, 2004}

\newsec{Introduction}

In the last couple of years, field theories defined on noncommutative 
spacetimes have been explored extensively, mainly due to their realization
in string theory. In particular, theories on spacetimes endowed with Moyal 
type deformations have been discussed. Besides purely bosonic deformations, 
also deformed superspaces became of interest (see, e.g., references 
\refs{\BrinkNB,\FerraraMM,\KosinskiXU,\KlemmYU,\BuchbinderAT} and more recent 
ones \refs{\FerraraXY,\IvanovTE,\FerraraXK}). As it was shown in 
\refs{\OoguriQP,\deBoerDN,\OoguriTT,\SeibergYZ,\BerkovitsKJ},
deformed superspaces arise quite naturally in string theory, as well.
For that reason, it is important to study such spaces as well as theories
defined on them. Within the past few months, several authors have
dealt with, for instance, deformed versions of the Wess-Zumino model and super
Yang-Mills theory. Also quantum aspects 
\refs{\BrittoAJ,\BrittoAJj,\GrisaruFD,\BrittoKG,\LuninBM,\BerensteinSR,
\AlishahihaKG} and nonperturbative 
solutions, such as instantons \refs{\ImaanpurJJ,\GrassiQK,\BrittoUV,\ArakiMQ} 
have been explored.

However, only deformed $\CN\leq 2$ super Yang-Mills theories have 
been discussed in the literature. This is probably due to the lack 
of a proper superspace formulation of the actions for $\CN=3,4$ 
super Yang-Mills theory -- even in the undeformed case. A loophole 
to this obstruction is to consider the constraint equations 
instead of an action. In the undeformed setup, it was pointed out 
in \WittenNT\ and proven in \refs{\HarnadVK,\HarnadBC} that there 
is a one-to-one correspondence between the equations of motion and 
the aforementioned constraint equations. The latter ones are 
defined on the superspace $\IR^{(4|4\CN)}$ and amount to a 
flatness condition on the superconnection\foot{ Note that even 
after a successful deformation of the equations of motion, one 
still had to find the corresponding action (in component fields) 
for a full description.}. 

In this paper we are going to use this fact to derive the equations of motion
of deformed super Yang-Mills theory by starting from properly deformed 
constraint equations. Since $\CN=3$ and $\CN=4$ super Yang-Mills theory are 
basically equivalent, we may just consider the $\CN=4$ case. In 
deriving the superfield expansions, we propose a generalization of
the Seiberg-Witten map \SeibergVS\ to superspace. For simplicity
we shall restrict ourselves to first order in the deformation.

The paper is organized as follows. In section $2$, we begin with a brief review
on $\CN=4$ super Yang-Mills theory which is defined on four-dimensional 
Euclidean spacetime. In section $3$, we then introduce the deformed
superspace $\IR^{(4|4\CN)}_\hbar$. Having fixed the setup, we start in section 
$4$ from the deformed constraint equations and derive the deformed equations
of motion by using the abovementioned Seiberg-Witten map.
Finally, in the appendix A we briefly review the expansion of the undeformed
superfields.

\newsec{$\CN=4$ super Yang-Mills theory}

\subsec{Generalities} 

We begin our considerations by fixing our notation and conventions, which to 
large extent coincide with those of \refs{\WessCP}. First of all, we shall 
always make the identification
$$ x^\mu\sim x^{\a\ad}, $$
where $\a,\b,\ldots,\ad,\bd,\ldots = 1,2$. Isospin indices will be denoted by
small Latin letters starting from the middle of the alphabet, i.e., 
$i,j,\ldots=1,\ldots,4$. Moreover, we use
\eqn\epsdef{ 
  (\epsilon^{\a\b})=(\epsilon^{\ad\bd})=\left(\matrix{0&1\cr -1&0\cr}\right)
   \qquad{\rm and}\qquad
  (\epsilon_{\a\b})=(\epsilon_{\ad\bd})=\left(\matrix{0&-1\cr 1&0\cr}\right),
}
with $\epsilon_{\a\d}\epsilon^{\d\b}=\delta^\b_\a$ and 
$\epsilon_{\ad\dd}\epsilon^{\dd\bd}=\delta^\bd_\ad.$ Spinors with upper and 
lower indices are related via the $\epsilon$-tensors, i.e.,
\eqn\spinconi{\psi^{i\a}=\epsilon^{\a\b}\psi_\b^i\qquad{\rm and}\qquad
             \psi_\a^i=\epsilon_{\a\b}\psi^{i\b}.}
Similarly, we have for the dotted ones
\eqn\spinconii{{\bar\psi}_i^\ad=\epsilon^{\ad\bd}{\bar\psi}_{i\bd}
               \qquad{\rm and}\qquad
              {\bar\psi}_{i\ad}=\epsilon_{\ad\bd}{\bar\psi}^\bd_i.}
It is important to stress that on Euclidean space there is no relation between
undotted and dotted spinors. Throughout this paper we use the following spinor 
summation convention:
\eqn\spinconiii{\eqalign{
                \psi\chi=\psi^{i\a}\chi^i_\a=-\psi^i_\a\chi^{i\a}=\chi^{i\a}
                \psi^i_\a=\chi\psi,\cr
                {\bar\psi}\bchi={\bar\psi}_{i\ad}\bchi^\ad_i=-{\bar\psi}^\ad_i
                \bchi_{i\ad}=\bchi_{i\ad}{\bar\psi}^\ad_i=\bchi{\bar\psi}.\cr
}}

\subsec{Euclidean $\CN=4$ super Yang-Mills action}

To write down the super Yang-Mills action, we recall first that the
automorphism group of $\CN=4$ supersymmetry on four-dimensional Euclidean space
is $SO(5,1)$. The $\CN=4$ supermultiplet consists of a gauge field $A_{\a\ad}$,
six scalars $W_{ij}=-W_{ji}$ and eight Weyl fermions $\chi^i_\a$ and 
$\bchi_{i\ad}$. All of these fields are subject to a specific reality 
condition induced by the anti-linear involutive automorphism 
$\sigma$:\foot{Summation over repeated indices is implied.}${}^,$\foot{Recall 
that a field $f$ is said to be real if it is a fixed point of the involution
$\sigma$, i.e., $\sigma(f)=f$.} 
\eqna\rc
$$\eqalignno{\sigma(A_{\a\bd})    & = -\epsilon_{\a\b}\epsilon_{\bd\gd}
                                      (A_{\b\gd})^\dagger,
                                  &   \rc a\cr
             \sigma(W_{ij})       & = -T^k_i (W_{kl})^\dagger T^l_j, &\rc b\cr
             \sigma(\chi^i_\a)    & = \epsilon_{\a\b}T^i_j(\chi^j_\b)^\dagger, 
                                  &   \rc c\cr
             \sigma(\bchi_{i\ad}) & = \epsilon_{\ad\bd}T^j_i
                                      (\bchi_{j\bd})^\dagger.
                                  &   \rc d\cr
}$$ 
The matrix $(T^i_j)$ is given by the following expression:
$$\eqalignno{(T^i_j)& = \left(\matrix{0  & 1 & 0  & 0 \cr
                                   -1  & 0 & 0  & 0 \cr
                                    0  & 0 & 0  & 1 \cr
                                    0  & 0 & -1 & 0 \cr}\right). 
                    &   \rc e\cr
}$$

{}For a detailed review on supersymmetry, especially on Euclidean spaces we
refer the reader to reference \VanProeyenNI. Note that all of the above fields
live in the adjoint representation of some compact gauge group $G$.
 
Using the conventions given in the previous subsection, the $\CN=4$ super 
Yang-Mills action on $\IR^4$ takes the following form
\eqn\syma{\eqalign{
          S=\int d^4x\ {\tr}\left\{-{\sfrac{1}{2}}\epsilon^{\a\b}
          \epsilon^{\ad\bd}\right.&
          \left.(\nabla_{\a\ad}W^{ij})(\nabla_{\b\bd}W_{ij})
          +\epsilon^{\ad\gd}\epsilon^{\bd\dd}f_{\ad\bd}f_{\gd\dd}+
          \epsilon^{\a\g}\epsilon^{\b\d}f_{\a\b}f_{\g\d}\right.\cr
          &\left.-\ {\sfrac{1}{2}}\epsilon_{ijkl}
          \epsilon^{\a\b}\chi^k_\a
          [\chi^l_\b,W^{ij}]-\epsilon^{\ad\bd}\bchi_{i\ad}[\bchi_{j\bd},W^{ij}]
          \right.\cr
          &\kern-1.5cm\left.+\ {\sfrac{1}{8}}[W^{ij},W^{kl}][W_{ij},W_{kl}]
          +\epsilon^{\a\b}\epsilon^{\bd\gd}(\chi^i_\a(\nabla_{\b\bd}
          \bchi_{i\gd})-(\nabla_{\b\bd}\chi^i_\a)\bchi_{i\gd})
          \right\},\cr            
}}
where we have abbreviated
$$ W^{ij}\equiv{\sfrac{1}{2}}\epsilon^{ijkl}W_{kl}.$$
Moreover, the bosonic curvature is decomposed (in self-dual and 
anti-self-dual parts) as
\eqn\decompboscurv{[\nabla_{\a\ad},\nabla_{\b\bd}]=
                \epsilon_{\ad\bd}f_{\a\b}+\epsilon_{\a\b}f_{\ad\bd}.}
The equations of motion induced by the action \syma\ read as
\eqna\diraceq
$$\eqalignno{\epsilon^{\a\b}\nabla_{\a\ad}\chi^i_\b+{\sfrac{1}{2}}
             \epsilon^{ijkl}[W_{kl},\bchi_{j\ad}] & = 0, &\diraceq a\cr
             \epsilon^{\ad\bd}\nabla_{\a\ad}\bchi_{i\bd}+[W_{ij},\chi^j_\a]
                                                  & = 0 &\diraceq b\cr
}$$
and
\eqna\curveomeq
$$\eqalignno{\epsilon^{\ad\bd}\nabla_{\g\ad}f_{\bd\gd}+\epsilon^{\a\b}
             \nabla_{\a\gd}f_{\b\g}&={\sfrac{1}{4}}\epsilon^{ijkl}
             [\nabla_{\g\gd}W_{ij},W_{kl}] +\{\chi^i_\g,\bchi_{i\gd}\} 
                                   , &\curveomeq a\cr
             \kern-1cm\epsilon^{\a\b}\epsilon^{\ad\bd}\nabla_{\a\ad}
             \nabla_{\b\bd}W_{ij}
             -{\sfrac{1}{4}}\epsilon^{klmn}[W_{mn}&,[W_{kl},W_{ij}]]
             &\cr 
             & ={\sfrac{1}{2}}\epsilon_{ijkl}\epsilon^{\a\b}
             \{\chi^k_\a,\chi^l_\b\}+\epsilon^{\ad\bd}\{\bchi_{i\ad},
             \bchi_{j\bd}\}        . &\curveomeq b\cr
}$$
The action is invariant under the following supersymmetry transformations
\eqna\suptransrules
$$\eqalignno{\d_{\xi,{\bar\xi}} 
              A_{\a\ad}\ =   & \ -\epsilon_{\a\b}\xi^{i\b}\bchi_{i\ad}+
             \epsilon_{\ad\bd}{\bar\xi}^\bd_i\chi^i_\a,
                                & \suptransrules a\cr
             \d_{\xi,{\bar\xi}} W_{ij}\ = & \ \epsilon_{ijkl}\xi^{k\a}\chi^l_\a
             -{\bar\xi}^\ad_i\bchi_{j\ad}+{\bar\xi}^\ad_j\bchi_{i\ad},
                                & \suptransrules b\cr
             \d_{\xi,{\bar\xi}}
             \chi^i_\a\ =    & \ -2\xi^{i\b}f_{\a\b}+{\sfrac{1}{2}}
             \epsilon_{\a\b}\epsilon^{iklm}\xi^{j\b}[W_{lm},W_{jk}]-
             \epsilon^{ijkl}{\bar\xi}^\ad_j\nabla_{\a\ad}W_{kl}, 
                                & \suptransrules c\cr
             \d_{\xi,{\bar\xi}}
             \bchi_{i\ad}\ = & \ 2\xi^{j\a}\nabla_{\a\ad}W_{ij}+
             2{\bar\xi}^\bd_i f_{\ad\bd}+{\sfrac{1}{2}}\epsilon_{\ad\bd}
             \epsilon^{jklm}{\bar\xi}^\bd_j[W_{lm},W_{ik}],
                                & \suptransrules d\cr
}$$
where $\xi^{i\a}$ and ${\bar\xi}^\ad_i$ are constant Weyl spinors with 
$$\eqalignno{\sigma(\xi^{i\a})=\epsilon^{\a\b}T^i_j(\xi^{j\b})^*
                    &\qquad{\rm and}\qquad\sigma({\bar\xi}^\ad_i)=
                    \epsilon^{\ad\bd}T^j_i({\bar\xi}^\bd_j)^*, &\rc f\cr
}$$
which is an immediate consequence of \rc{c,d}. Here, ``$*$'' denotes complex 
conjugation.

\newsec{Deformed superspace $\IR^{(4|4\CN)}_\hbar$}

\subsec{Definition of $\IR^{(4|4\CN)}$}

Before we start to discuss deformed superspaces, we must say a few words
about graded Poisson structures. Let $\V$ be a vector space over $\IR$ or 
$\IC$ equipped with a $\IZ_2$-grading, i.e., $\V$ is decomposed as 
$\V\cong\bigoplus_{p=0,1}\V_p$. Elements of $\V_0$ and $\V_1$ are said 
to have even ($p=0$) and odd ($p=1$) parity, respectively. Moreover, we can 
lift the vector space $\V$ to a graded 
algebra by endowing it with an associative product which is assumed to respect 
the grading, i.e., for $f,g\in\V$ and $\cdot\,:\,f\otimes g\mapsto f\cdot g$ we
write
$$ p(f\cdot g)=p(f)+p(g)\ \ {\rm mod}\ \ 2. $$ 
By introducing a graded Lie bracket
$[\,,\,\}\,:\,\V\otimes\V\to\V$ defined by
\eqn\supercom{[f,g\}\equiv f\cdot g-(-)^{p_fp_g}g\cdot f,}
the graded algebra $\V$ becomes a graded Lie algebra. In the following
we shall omit ``$\cdot$''. Of course, \supercom\ satisfies a graded Jacobi 
identity,
\eqn\jacobi{[f,[g,h\}\}+(-)^{p_f(p_g+p_h)}[g,[h,f\}\}+
            (-)^{p_h(p_f+p_g)}[h,[g,f\}\}=0,}
for any $f,g,h\in{\frak V}$. Then we define the superspace $\IR^{(4|4\CN)}$ by
\eqn\defss{\IR^{(4|4\CN)}\equiv C^\infty(\IR^4)\otimes
           \Lambda^\bullet(\IR^{4\CN}),}
where $\Lambda^\bullet(\IR^{4\CN})$ denotes the exterior algebra of 
$\IR^{4\CN}$. The elements of $\IR^{(4|4\CN)}$ are called superfields.
The algebra $\IR^{(4|4\CN)}$ is finitely generated by
$(X^a)=(x^{\a\ad},\t^{i\a},\bt^\ad_i)$ with $\a,\b,\ldots,\ad,\bd,\ldots=1,2$ 
and $i,j,\ldots=1,\ldots,\CN$. Indices $a,b,\ldots$ represent all generators.
To these generators\foot{In the following, we shall call them loosely 
``coordinates''.} we assign the following parities:
$$ p_{x^{\a\ad}}\equiv 0\qquad{\rm and}\qquad 
   p_{\t^{i\a}}\equiv p_{\bt^\ad_i}\equiv 1.$$
The action of $\sigma$ is given by
$$\eqalignno{\kern-1cm\sigma(x^{\a\bd})&=\epsilon^{\a\b}(x^{\b\gd})^*
             \epsilon^{\bd\gd},\quad
             \sigma(\t^{i\a})=\epsilon^{\a\b}T^i_j(\t^{j\b})^*
             \quad{\rm and}\quad
             \sigma(\bt^\ad_i)=\epsilon^{\ad\bd}T^j_i(\bt^\bd_j)^*,
             & \rc g\cr
}$$
where $T^i_j$ is given by \rc{e}.
Any superfield $f\in\IR^{(4|4\CN)}$ may be expanded in terms of the $\t^{i\a}$ 
and $\bt^\ad_i$ coordinates as
\eqn\expan{f=\nothiii{f}(x)\ \ +\sum_{0<|I|,|{\bar J}|\leq 2\CN} 
             f_{I,{\bar{J}}}(x)\ \t^I\bt^{\bar J},}
where $I$ and ${\bar J}$ are multiindices with $I=(i_1\a_1,\ldots,i_{|I|}
\a_{|I|})$ and 
${\bar J}=(\sm{\ad_1 }{i_1\ },\ldots,\sm{\ad_{|{\bar J}|}}{i_{|{\bar J}|}})$.
In \expan\ we suppressed the wedge symbol. 

On $\IR^{(4|4\CN)}$ we may introduce left and right derivations. A left 
derivation is a linear map $\overrightarrow{\partial}$ which satisfies
$$ \overrightarrow{\partial}(fg)=\overrightarrow{\partial}(f)g
   +(-)^{p_\partial p_f}f\overrightarrow{\partial}(g) $$
for $f,g\in\IR^{(4|4\CN)}$. Similarly, we may introduce right derivations as
$$ f\overleftarrow{\partial}\equiv (-)^{p_\partial(p_f+1)}
    \overrightarrow{\partial}f. $$
In these equations $p_\partial$ is called the degree of the derivation.

A graded (or super) Poisson structure on $\IR^{(4|4\CN)}$ is a graded Lie
algebra structure \supercom, which satisfies the Jacobi identity \jacobi\ and
\eqn\supercomii{\eqalign{
                [f,gh\}& =[f,g\}h+(-)^{p_fp_g}g[f,h\},\cr
                [fg,h\}& =f[g,h\}+(-)^{p_gp_h}[f,h\}g,\cr
}}
for $f,g,h\in\IR^{(4|4\CN)}$.

\subsec{Superderivatives, supercharges and supersymmetry algebra}

Let $f\in\IR^{(4|4\CN)}$. The left superderivatives are defined in the usual 
way as
\eqn\superder{\eqalign{
              \overrightarrow{D}_{i\a}f&\equiv\overrightarrow{\partial}_{i\a}f
              +\bt^\ad_i\partial_{\a\ad}f,\cr
              \overrightarrow{{\bar D}}\ \!\!^i_\ad f&\equiv
              -\overrightarrow{\bar\partial}\ \!\!^i_\ad f
              -\t^{i\a}\partial_{\a\ad}f.\cr
}}
They satisfy the following algebra:
\eqn\superalgi{\{\overrightarrow{D}_{i\a},\overrightarrow{D}_{j\b}\}=0,\qquad
               \{\overrightarrow{{\bar D}}\ \!\!^i_\ad,
                 \overrightarrow{{\bar D}}\ \!\!^j_\bd\}=0
                 \qquad{\rm and}\qquad
               \{\overrightarrow{D}_{i\a},
                 \overrightarrow{{\bar D}}\ \!\!^j_\bd\}=-2\delta^i_j
                \partial_{\a\bd}.}
The definition of the right superderivatives $\overleftarrow{D}_{i\a}$ and 
$\overleftarrow{{\bar D}}\ \!\!^i_\ad$ is then immediate. 

The left supercharge operators are given by the expressions
\eqn\supercharge{\eqalign{
              \overrightarrow{Q}_{i\a}f&\equiv\overrightarrow{\partial}_{i\a}f
              -\bt^\ad_i\partial_{\a\ad}f,\cr
              \overrightarrow{{\bar Q}}\ \!\!^i_\ad f&\equiv
              -\overrightarrow{\bar\partial}\ \!\!^i_\ad f
              +\t^{i\a}\partial_{\a\ad}f.\cr
}}
They obviously satisfy
\eqn\superalgii{\{\overrightarrow{Q}_{i\a},\overrightarrow{Q}_{j\b}\}=0,\qquad
               \{\overrightarrow{{\bar Q}}\ \!\!^i_\ad,
                 \overrightarrow{{\bar Q}}\ \!\!^j_\bd\}=0
                 \qquad{\rm and}\qquad
               \{\overrightarrow{Q}_{i\a},
                 \overrightarrow{{\bar Q}}\ \!\!^j_\bd\}=2\delta^i_j
                \partial_{\a\bd}}
and they anticommute with the left superderivatives \superder. Recall that
the supercharges \supercharge\ generate the following (super)translations on 
the superspace
\eqn\supertrans{\eqalign{
                \t^{i\a}\mapsto\t^{\prime\,i\a}   & = \t^{i\a}+
                \xi^{i\a},\qquad
                \bt^\ad_i\mapsto\bt^{\prime\ad}_i 
                = \bt^\ad_i+{\bar\xi}^\ad_i,\cr
                x^{\a\ad}\mapsto x^{\prime\a\ad} & = x^{\a\ad}+a^{\a\ad}
                -(\xi^{i\a}{\bt}^\ad_i+{\bar\xi}^\ad_i\t^{i\a}),
}}
where $\xi^{i\a}$ and ${\bar\xi}^\ad_i$ are constant Majorana-Weyl spinors and 
$a^{\a\ad}$ represents a constant four-vector.

\subsec{Definition of $\IR^{(4|4\CN)}_\hbar$}

Having fixed the undeformed setup we are now ready for the discussion of 
a non(anti)commutative extension of our theory. 

Let $[\,,\,\}$ be a super Poisson structure. Moreover, we define  
an operator $P\,:\,\IR^{(4|4\CN)}\otimes \IR^{(4|4\CN)}\to \IR^{(4|4\CN)}$ by
$$ P(f\otimes g)\equiv[f,g\}.$$ 
Now we consider the algebra $\IR^{(4|4\CN)}[[\hbar]]$ which we abbreviate
by $\IR^{(4|4\CN)}_\hbar$ in the sequel. The variable $\hbar$ is some 
parameter in which we consider a formal power series expansion. Of
course, in the limit $\hbar\to 0$ we recover $\IR^{(4|4\CN)}$. Then we define
a star product by
\eqn\defstar{\eqalign{
             \star\,:\,\IR^{(4|4\CN)}_\hbar\otimes\IR^{(4|4\CN)}_\hbar &\ \to\
              \IR^{(4|4\CN)}_\hbar,\cr
             f\otimes g \ \mapsto\ e^{\hbar P}(f\otimes g)=\sum_n{\hbar^n
             \over n!}P^n(f\otimes g)&\ =\ fg+\hbar [f,g\}+\CO(\hbar^2).
}}
In this paper we assume that $P$ is a bi-differential operator of the 
form
\eqn\defp{P^n(f\otimes g)={\sfrac{1}{2^n}}C^{i_1\a_1,j_1\b_1}\cdots
                          C^{i_n\a_n,j_n\b_n}
                          f \overleftarrow{Q}_{i_1\a_1}\cdots
                            \overleftarrow{Q}_{i_n\a_n}
                        \overrightarrow{Q}_{j_n\b_n}\cdots
                        \overrightarrow{Q}_{j_1\b_1} g,}
where $C^{i\a,j\b}\in{\rm Sym}(2\CN,\IC)$, i.e., $i,j=1,\ldots,\CN$. 
Moreover, $C^{i\a,j\b}$ is assumed to be constant. Note that in this way we 
have defined an associative and nilpotent star product, i.e., the series 
expansion in \defstar\ is finite and goes up to $\CO(\hbar^{2\CN})$. 

Equation \defp\ implies that
\eqn\defstaronfunct{f\star g=f\, \exp\left\{{\sfrac{\hbar}{2}}
   \overleftarrow{Q}_{i\a}C^{i\a,j\b}
   \overrightarrow{Q}_{j\b}\right\}\, g.}
Using the definition \superder, one may readily verify that
\eqn\coordstar{\eqalign{
               x^{\a\ad}\star x^{\b\bd}&= x^{\a\ad}x^{\b\bd}-{\sfrac{\hbar}{2}}
                                          C^{i\a,j\b}\bt^\ad_i\bt^\bd_j,\cr
               x^{\a\ad}\star\t^{j\b}&=  x^{\a\ad}\t^{j\b}+{\sfrac{\hbar}{2}}
                                          C^{i\a,j\b}\bt^\ad_i,\cr
               \t^{i\a}\star\t^{j\b} &=\t^{i\a}\t^{j\b}+{\sfrac{\hbar}{2}}
                                        C^{i\a,j\b}.\cr
}}
Therefore we may introduce a star supercommutator by
\eqn\starsupercom{[f,g\}_\star\equiv f\star g-(-)^{p_fp_g}g\star f}
leading to
\eqn\deformi{\eqalign{
             [x^{\a\ad},x^{\b\bd}]_\star &\ =\ -\hbar\,
                                            C^{i\a,j\b}\bt^\ad_i\bt^\bd_j,\cr
             [x^{\a\ad},\t^{j\b}]_\star &\ =\ \hbar\,
                                            C^{i\a,j\b}\bt^\ad_i,\cr
             \{\t^{i\a},\t^{j\b}\}_\star &\ =\ \hbar\,
                                            C^{i\a,j\b}.\cr
}}
Of course, now we also have a star Jacobi identity
\eqn\jacobinc{[f,[g,h\}_\star\}_\star+(-)^{p_f(p_g+p_h)}
              [g,[h,f\}_\star\}_\star+
              (-)^{p_h(p_f+p_g)}[h,[g,f\}_\star\}_\star=0,}
for $f,g,h\in\IR^{(4|4\CN)}_\hbar$.

Obviously, the algebra 
\deformi\ does not transform covariantly under the full set of
(super)translations \supertrans. Depending on the rank of the deformation
matrix $C^{i\a,j\b}$, the supersymmetry will partially be broken.
Note that here we are assuming an 
undeformed parameter algebra, i.e., $\epsilon^{i\a}$, $\ebar^\ad_i$ and 
$a^{\a\ad}$ are kept (anti)commuting.

\subsec{Deformed supersymmetry algebra}

As it will become clear momentarily, it will be convenient for us to use chiral
coordinates on $\IR^{(4|4\CN)}_\hbar$ instead of 
$(X^a)=(x^{\a\ad},\t^{i\a},\bt^\ad_i)$. These are defined by\foot{Note that
$\t^{i\a}\star\bt_i^\ad=\t^{i\a}\bt_i^\ad$.}
\eqn\chicor{(X^a)= (x^{\a\ad},\t^{i\a},\bt^\ad_i)\mapsto
           (Y^a) = (y^{\a\ad}=x^{\a\ad}+\t^{i\a}\bt_i^\ad,\t^{i\a},\bt^\ad_i).}
It is easy to check that the involution $\sigma$ acts as
$$\eqalignno{\sigma(y^{\a\bd})&=\epsilon^{\a\b}(y^{\b\gd})^*
             \epsilon^{\bd\gd}.&\rc i
}$$
In the coordinates \chicor\ the superderivatives \superder\ and the 
supercharges \supercharge\ take the following form
\eqna\superderchar
$$\eqalignno{\overrightarrow{D}_{i\a}=\overrightarrow{\partial}_{i\a}+
             2\bt^\ad_i\partial_{\a\ad}\qquad & {\rm and}\qquad
             \overrightarrow{\bar{D}}\ \!\!^i_\ad
             = -\overrightarrow{\bar{\partial}}\ \!\!^i_\ad,
                                              & \superderchar a\cr
             \overrightarrow{Q}_{i\a}=\overrightarrow{\partial}_{i\a}
                                       \qquad & {\rm and}\qquad
             \overrightarrow{\bar{Q}}\ \!\!^i_\ad=
             -\overrightarrow{{\bar\partial}}\ \!\!^i_\ad+
             2\t^{i\a}\partial_{\a\ad},       & \superderchar b\cr
}$$
where now the $\partial_{\a\ad}$s are partial derivatives with respect to
$y^{\a\ad}$. 

One may readily check that the (anti)commutation relations \deformi\ become
\eqn\deformii{\eqalign{
             [y^{\a\ad},y^{\b\bd}]_\star &\ =\ 0,\cr
             [y^{\a\ad},\t^{j\b}]_\star &\ =\ 0,\cr
             \{\t^{i\a},\t^{j\b}\}_\star &\ =\ \hbar\,
                                            C^{i\a,j\b}.\cr
}}

{}Furthermore, it is obvious from the explicit form \superderchar{a} of the 
superderivatives that they satisfy
\eqn\superalgncI{\eqalign{
                 \{\overrightarrow{D}_{i\a},\overrightarrow{D}_{j\b}\}_\star
                  & = 0,\cr
                 \{\overrightarrow{{\bar D}}\ \!\!^i_\ad,
                 \overrightarrow{{\bar D}}\ \!\!^j_\bd\}_\star&=0,\cr
                 \{\overrightarrow{D}_{i\a},
                 \overrightarrow{{\bar D}}\ \!\!^j_\bd\}_\star&=-2\delta^i_j
                 \partial_{\a\bd},\cr
}}
i.e., the same algebra as in the undeformed case. The supercharge operators
are subject to the following relations
\eqn\superalgncII{\eqalign{
                  \{\overrightarrow{Q}_{i\a},\overrightarrow{Q}_{j\b}\}_\star
                   & = 0,\cr
                  \{\overrightarrow{{\bar Q}}\ \!\!^i_\ad,
                  \overrightarrow{{\bar Q}}\ \!\!^j_\bd\}_\star
                  &=  4\hbar\,C^{i\a,j\b}\partial_{\a\ad}\partial_{\b\bd},\cr
                  \{\overrightarrow{Q}_{i\a},
                  \overrightarrow{{\bar Q}}\ \!\!^j_\bd\}_\star&=2\delta^i_j
                 \partial_{\a\bd},\cr
}}
while the star anticommutators between the superderivatives and supercharges
do still vanish. In the sequel, we shall refer to \superalgncI\ and
\superalgncII\ as the deformed supersymmetry algebra. The explicit form
of algebra \superalgncII\ makes the supersymmetry breaking apparent.
Note that the change
of coordinates \chicor\ was needed since otherwise the 
$\overrightarrow{{\bar D}}\ \!\!^i_\ad$s would not be derivations with respect 
to the star product. This property, however, is essential in the subsequent 
discussion.

\newsec{Constraint equations and field equations on $\IR^{(4|16)}_\hbar$}

Now we are discussing the $\CN=4$ case. The goal of this section is to derive 
the field equations of deformed $\CN=4$ super Yang-Mills theory from
the constraint equations. The discussion in the undeformed case is given in
\refs{\HarnadVK,\HarnadBC}. 

In the sequel, let ${\frak g}$ be the gauge algebra. For instance, we could 
take ${\frak g}$ to be ${\frak u}(n)$. Generally speaking, ${\frak g}$ is 
some enveloping algebra. In order to simplify notation, we shall omit the 
arrows over the derivatives. If not stated differently, all derivatives are 
left derivatives.

\subsec{Superfield equations from constraint equations}

{}From now on we shall adopt the common convention and write always 
hats when we mean deformed superfields. 

Starting point is the deformed constraint equations 
\eqna\constraintnc
$$\eqalignno{
 \{\hnab_{i\a},\hnab_{j\b}\}_\star   & = -2\epsilon_{\a\b}{\hat W}_{ij},
                                     &   \constraintnc a\cr
 \{\hbnab\ \!\!^i_\ad,\hbnab\ \!\!^j_\bd\}_\star 
                                     & = -\epsilon_{\ad\bd}\epsilon^{ijkl} 
                                         {\hat W}_{kl},
                                     &   \constraintnc b\cr
 \{\hnab_{i\a},\hbnab\ \!\!^j_\bd\}_\star  & =-2\delta_i^j\ \hnab_{\a\bd}.
                                     &   \constraintnc c\cr
}$$ 
The covariant derivatives in \constraintnc{} are given by
\eqn\covdernc{\hnab_{i\a} = D_{i\a}+[{\hat\o}_{i\a},\,\,\}_\star,\quad
              \hbnab\ \!\!^i_\ad =\Db^i_\ad-[{\hat\bo}\ \!\!^i_\ad,\,\,\}_\star
              \quad{\rm and}\quad
              \hnab_{\a\bd}=\partial_{\a\bd}+[{\hat A}_{\a\bd},\,\,\,]_\star.}
More explicitly, the equations \constraintnc{} read as
\eqna\constraintnci
$$\eqalignno{
 D_{i\a}{\hat\o}_{j\b}+D_{j\b}{\hat\o}_{i\a}+\{{\hat\o}_{i\a},
                               {\hat\o}_{j\b}\}_\star
                              & = -2\epsilon_{\a\b}{\hat W}_{ij},
                              &   \constraintnci a\cr
 \Db^i_\ad{\hat\bo}\ \!\!^j_\bd+\Db^j_\bd{\hat\bo}\ \!\!^i_\ad-
                      \{{\hat\bo}\ \!\!^i_\ad,{\hat\bo}\ \!\!^j_\bd\}_\star 
                              & = \epsilon_{\ad\bd}\epsilon^{ijkl} 
                                  {\hat W}_{kl},
                              &  \constraintnci b\cr
 D_{i\a}{\hat\bo}\ \!\!^j_\bd-\Db^j_\bd{\hat\o}_{i\a}+\{{\hat\o}_{i\a},
                 {\hat\bo}\ \!\!^j_\bd\}_\star
                              & = 2\delta_i^j\ {\hat A}_{\a\bd}.
                              &  \constraintnci c\cr
}$$  

As usual, we decompose the bosonic curvature as
\eqn\bossupcurv{
   [\hnab_{\a\ad},\hnab_{\b\bd}]_\star=
        \epsilon_{\ad\bd}{\hat{f}}_{\a\b}+\epsilon_{\a\b}{\hat{f}}_{\ad\bd}.}
Then we define the superspinor fields by the equations
\eqna\superspinor   
$$\eqalignno{[\hnab_{i\a},\hnab_{\b\bd}]_\star
                            \equiv\ &\epsilon_{\a\b}{\hat\bchi}_{i\bd},
                                              &\superspinor a\cr
             [\hbnab\ \!\!^i_\ad,\hnab_{\b\bd}]_\star
                            \equiv\ &\epsilon_{\ad\bd}{\hat\chi}\ \!\!^i_\b,
                                              &\superspinor b\cr
}$$
which are
\eqna\superspinori  
$$\eqalignno{D_{i\a}{\hat A}_{\b\bd}-\partial_{\b\bd}{\hat\o}_{i\a}+
             [{\hat\o}_{i\a},{\hat A}_{\b\bd}]_\star\ =\ 
              &\epsilon_{\a\b}{\hat\bchi}_{i\bd},
                                          &\superspinori a\cr
             \Db^i_\ad {\hat A}_{\b\bd}+\partial_{\b\bd}{\hat\bo}\ \!\!^i_\ad-
             [{\hat\bo}\ \!\!^i_\ad,{\hat A}_{\b\bd}]_\star\ =\ 
                          &\epsilon_{\ad\bd}{\hat\chi}\ \!\!^i_\b.
                                          &\superspinori b\cr
}$$
As an immediate consequence of the graded Bianchi identities,
$$[\hnab_a,[\hnab_b,\hnab_c\}_\star\}_\star+
             (-)^{p_a(p_b+p_c)}[\hnab_b,[\hnab_c,\hnab_a\}_\star\}_\star+
             (-)^{p_c(p_a+p_b)}[\hnab_c,[\hnab_a,\hnab_b\}_\star\}_\star=0,
$$
where $\hnab_a$ is either $\hnab_{\a\ad}$, $\hnab_{i\a}$ or 
$\hbnab\ \!\!^i_\ad$, we have
\eqna\supereqI  
$$\eqalignno{\hnab_{i\a}{\hat W}_{jk}=\           &
             \epsilon_{ijkl}{\hat\chi}\ \!\!^l_\a,      &\supereqI a\cr
             \hbnab\ \!\!^i_\ad {\hat W}_{jk}=\         &
             \delta^i_j{\hat\bchi}_{k\ad}-\delta^i_k{\hat\bchi}_{j\ad},
                                            &\supereqI b\cr
             \hnab_{i\a}{\hat\bchi}_{j\ad}=\     &
             -2\hnab_{\a\ad}{\hat W}_{ij},        &\supereqI c\cr
             \hbnab\ \!\!_\ad^{i}{\hat\chi}\ \!\!^{j}_\a=\   &
             -\epsilon^{ijkl}\hnab_{\a\ad}{\hat W}_{kl}.
                                            &\supereqI d\cr 
}$$
Considering the Bianchi identities for 
$(\hnab_{i\a},\hbnab\ \!\!^j_\bd,\hnab_{\g\gd})$ together with the 
Bianchi identities for $(\hnab_{i\a},\hnab_{j\b},{\hat W}_{lk})$ and equation 
\bossupcurv, we discover
\eqna\supereqII
$$\eqalignno{\hnab_{i\a}{\hat\chi}\ \!\!^j_\b =\    & 
             -2\delta^j_i{\hat{f}}_{\a\b}-{\sfrac{1}{2}}
                   \epsilon_{\a\b}\epsilon^{jklm}
             [{\hat W}_{lm},{\hat W}_{ik}]_\star,            &\supereqII a\cr
             \hbnab\ \!\!^j_\ad{\hat\bchi}_{i\bd}=\ &
             -2\delta^j_i{\hat{f}}_{\ad\bd}+{\sfrac{1}{2}}\epsilon_{\ad\bd}
             \epsilon^{jklm}[{\hat W}_{lm},{\hat W}_{ik}]_\star.
                                         &\supereqII b\cr
}$$
Similarly, again by using Bianchi identities a straightforward computation 
yields
\eqna\supereqIII
$$\eqalignno{\hnab_{i\a}{\hat{f}}_{\b\g} =\     &
             -{\sfrac{1}{2}}\epsilon^{\bd\gd}(\epsilon_{\a\b}\hnab_{\g\bd}
             {\hat\bchi}_{i\gd}+\epsilon_{\a\g}
                          \hnab_{\b\bd}{\hat\bchi}_{i\gd}),
                                         &\supereqIII a\cr
             \hbnab\ \!\!^i_\ad {\hat{f}}_{\b\g} =\   &
             {\sfrac{1}{2}}(\hnab_{\b\ad}{\hat\chi}\ \!\!^i_\g+
                          \hnab_{\g\ad}{\hat\chi}\ \!\!^i_\b),
                                         &\supereqIII b\cr
             \hnab_{i\a}{\hat{f}}_{\bd\gd} =\   &
             {\sfrac{1}{2}}(\hnab_{\a\bd}{\hat\bchi}_{i\gd}+\hnab_{\a\gd}
             {\hat\bchi}_{i\bd}),              &\supereqIII c\cr
             \hbnab\ \!\!^i_\ad {\hat{f}}_{\bd\gd} =\ & 
             -{\sfrac{1}{2}}\epsilon^{\b\g}(\epsilon_{\ad\bd}\hnab_{\g\gd}
             {\hat\chi}\ \!\!^i_\b+\epsilon_{\ad\gd}
                                 \hnab_{\g\bd}{\hat\chi}\ \!\!^i_\b).
                                         &\supereqIII d\cr
}$$
Furthermore, from equation \constraintnc{c} we deduce
\eqn\supereqIV{\hnab_{\a\ad}=-{\sfrac{1}{8}}\{\hnab_{i\a},\hbnab\ \!\!^i_\ad
   \}_\star.}
Applying it to ${\hat\chi}\ \!\!^i_\b$ and ${\hat\bchi}_{i\bd}$ and using 
the equations \superspinor{}, \supereqI{} and \supereqII{} we obtain the super 
Dirac equations
\eqna\diracnc
$$\eqalignno{\epsilon^{\a\b}\hnab_{\a\ad}{\hat\chi}\ \!\!^i_\b+{\sfrac{1}{2}}
             \epsilon^{ijkl}[{\hat W}_{kl},{\hat\bchi}_{j\ad}]_\star 
             & = 0, &\diracnc a\cr
             \epsilon^{\ad\bd}\hnab_{\a\ad}{\hat\bchi}_{i\bd}
              +[{\hat W}_{ij},{\hat\chi}\ \!\!^j_\a]_\star
                                                  & = 0. &\diracnc b\cr
}$$
Note that they look formally the same as in the undeformed setup, but this
time all products are replaced by star products.

Applying $\hnab_{m\g}$ to \diracnc{a} and to \diracnc{b} and using the 
equations
\superspinor{},  \supereqI{} and \supereqII{} we get the superfield equations 
of motion for the gauge field and the scalar multiplet
\eqna\curveomnc
$$\eqalignno{\epsilon^{\ad\bd}\hnab_{\g\ad}{\hat{f}}_{\bd\gd}+\epsilon^{\a\b}
             \hnab_{\a\gd}{\hat{f}}_{\b\g}&={\sfrac{1}{4}}\epsilon^{ijkl}
             [\hnab_{\g\gd}{\hat W}_{ij},{\hat W}_{kl}]_\star 
             +\{{\hat\chi}\ \!\!^i_\g,{\hat\bchi}_{i\gd}\}_\star
                                   , &\curveomnc a\cr
             \kern-1cm\epsilon^{\a\b}\epsilon^{\ad\bd}\hnab_{\a\ad}
             \hnab_{\b\bd}{\hat W}_{ij}
             -{\sfrac{1}{4}}
             \epsilon^{klmn}[{\hat W}_{mn}&,[{\hat W}_{kl},
             {\hat W}_{ij}]_\star]_\star
             &\cr 
             & ={\sfrac{1}{2}}\epsilon_{ijkl}\epsilon^{\a\b}
             \{{\hat\chi}\ \!\!^k_\a,{\hat\chi}\ \!\!^l_\b\}_\star+
              \epsilon^{\ad\bd}\{{\hat\bchi}_{i\ad},
             {\hat\bchi}_{j\bd}\}_\star        . &\curveomnc b\cr
}$$

\subsec{Superfield expansions}

In the previous subsection, we have derived the superfield equations of motion 
from the constraint equations on $\IR^{(4|16)}_\hbar$. Now we are interested in
the superfield expansions. Of particular interest is, of course, the zeroth 
order components of \diracnc{} and \curveomnc{}, i.e., those terms which do not
contain the odd coordinates $\t^{i\a}$ and $\bt^\ad_i$.

The problem we are faced with is to find a proper way to construct the 
deformed superfields which, as we have already seen in \SeibergYZ, will
differ from their undeformed pendants. In the appendix A, we review their
construction in the undeformed case.

One possibility is to find some proper recursion operator as was done
in the undeformed setup (see references \refs{\HarnadVK,\HarnadBC}). Following
this approach, one encounters the difficulty that all theta orders of the 
superfields do mix, i.e., one discovers a coupled non-linear system of 
algebraic equations for the component fields, which one has to solve 
simultaneously for all the components. However, already in the undeformed setup
the superfield expansions become quite lengthy as one considers higher and 
higher powers in the odd coordinates. Therefore we suggest to follow another 
way which is based on a generalization of the Seiberg-Witten map \SeibergVS\ to
superspace.\foot{In the case of purely bosonic deformations, Seiberg-Witten 
maps for $\CN=1$ superfields have been discussed in 
\refs{\DayiJU,\MikulovicSQ}.} 
Eventually, this yields a systematic way to construct the deformed 
superfield equations order by order in the deformation $\hbar$. Such a
generalization of the Seiberg-Witten map seems quite natural and has been
conjectured throughout the literature \refs{\deBoerDN,\SeibergYZ}.
We shall add some remarks on this topic in the conclusions.

Remember that we have one {\it fundamental\/} field in our theory. For 
instance, we may regard the gauge potential ${\hat\o}_{i\a}$ as fundamental. 
This simply means that all other field expansions, i.e., those for
${\hat\bo}\ \!\!^i_\ad$, ${\hat A}_{\a\ad}$, 
${\hat W}_{ij}$, ${\hat\chi}\ \!\!^i_\a$ and ${\hat\bchi}_{i\ad}$
are completely determined through ${\hat\o}_{i\a}$ via the constraint
equations \constraintnc{} and the definitions \superspinor{} (in a certain 
gauge; see below).

As in ordinary noncommutative field theory, the starting point is then the
equation
\eqn\sswm{{\hat\o}_{i\a}(\o+\d_\l\o,\bo+\d_\l\bo)\ =\
         {\hat\o}_{i\a}(\o,\bo)+\d_{{\hat\Lambda}}{\hat\o}_{i\a}(\o,\bo).}
Recall that infinitesimal gauge transformations of the undeformed superfield
$\o_{i\a}$ is induced by an even superfield $\l$ via
\eqn\gti{\d_\l\o_{i\a}=D_{i\a}\l+[\o_{i\a},\l],}
whereas for the deformed superfield we have
\eqn\gtii{\d_{\hat\L}{\hat\o}_{i\a}=D_{i\a}{\hat\L}+
           [{\hat\o}_{i\a},{\hat\L}]_\star.}
Instead of directly solving \sswm{} for the gauge potential and the gauge
parameter, we may consider a consistency condition of two successive gauge 
transformations (see, e.g., \JurcoRQ\ and references therein). Take, for 
instance, some superfield ${\hat\psi}$ which transforms in the fundamental 
representation of the gauge group, i.e.,
\eqn\gtiii{\d_{\hat\L}{\hat\psi}=-{\hat\L}\star{\hat\psi}}
and hence
\eqn\gtiv{ 
   [\d_{\hat\L},\d_{\hat\Sigma}]{\hat\psi}=-[{\hat\L},{\hat\Sigma}]_\star\star
   {\hat\psi}=\d_{[{\hat\L},{\hat\Sigma}]_\star}{\hat\psi}.}
Following \JurcoRQ, we are looking for gauge transformations of the type
$$ {\hat\L}(\lambda,\o,\bo)\equiv {\hat\L}_\l(\o,\bo).$$
Therefore we restrict the transformation \gtiii\ to
\eqn\gtv{\d_\l{\hat\psi}=-{\hat\L}_\lambda(\o,\bo)\star{\hat\psi}.}
Thus, equation \gtiv\ translates into
\eqn\gtvi{\d_\l{\hat\L}_\sigma-\d_\sigma{\hat\L}_\l+
          [{\hat\L}_\l,{\hat\L}_\sigma]_\star={\hat\L}_{[\l,\sigma]}.}
Now we assume that it is possible to expand ${\hat\L}_\l$ in powers of $\hbar$,
i.e., we write
\eqn\swexpan{{\hat\L}_\l=\l+\hbar\,{\hat\L}^1_\l+\CO(\hbar^2).}
Expanding equation \gtvi\ in powers of $\hbar$ and substituting equation
\swexpan, we realize that to zeroth order it is trivially satisfied. To first 
order in $\hbar$ we obtain
$$ \d_\l\hL^1_\s-\d_\s\hL^1_\l+[\l,\hL^1_\s]+[\hL^1_\l,\s]-{\sfrac{1}{2}}
    C^{i\a,j\b}\,[\partial_{i\a}\l,\partial_{j\b}\s]=\hL^1_{[\l,\s]}.$$
A solution to this equation is of the form
\eqna\swsol
$$\eqalignno{\hL_\l^1 &=-{\sfrac{1}{4}}C^{i\a,j\b}[\partial_{i\a}\l,\O_{j\b}],
                      &\swsol a\cr
}$$
where we have introduced
\eqn\defbigo{\O_{i\a}\equiv\o_{i\a}+\bt^\bd_j({\bar D}^j_\bd\o_{i\a}+
                      D_{i\a}\bo^j_\bd+\{\bo^j_\bd,\o_{i\a}\}).}
In order to verify \swsol{a}, we note that infinitesimal gauge transformations 
act on $\O_{i\a}$ as 
$$\d_\l\O_{i\a}=\partial_{i\a}\l+[\O_{i\a},\l].$$
Therefore we may write
\eqn\swsolexp{\hL_\l=\l-{\sfrac{\hbar}{4}}C^{i\a,j\b}
              [\partial_{i\a}\l,\O_{j\b}]+\CO(\hbar^2).}

Having derived the expansion for the gauge parameter to first order in $\hbar$,
it is now easy to give the expansion for the super gauge potential
${\hat\o}_{i\a}$. Consider the expansion
\eqn\swexpani{{\hat\o}_{i\a}=\o_{i\a}+\hbar\,{\hat\o}^1_{i\a}
               +\CO(\hbar^2).}
Equation \gtii\ then yields
$$\d_\l{\hat\o}^1_{i\a}=D_{i\a}\hL^1_\l+[{\hat\o}^1_{i\a},\l]+[\o_{i\a},
                        \hL^1_\l]
                        +{\sfrac{1}{2}}C^{j\b,k\g}\{\partial_{j\b}\o_{i\a},
                       \partial_{k\g}\l\}.$$
Note that our equations are again satisfied identically to zeroth order. 
Substituting our solution \swsol{a} into the above equation, one finds after 
some algebraic manipulations that 
$$\eqalignno{{\hat\o}_{i\a}^1&={\sfrac{1}{4}}C^{j\b,k\g}
              \{\O_{j\b},\partial_{k\g}\o_{i\a}+R_{k\g,i\a}\},& \swsol b\cr
}$$
with
\eqn\defofr{R_{i\a,j\b}\equiv \partial_{i\a}\o_{j\b}+D_{j\b}\O_{i\a}+
            \{\o_{j\b},\O_{i\a}\}}
is a solution. 
Note that $R_{i\a,j\b}$ transforms under infinitesimal gauge transformations
as
$$\d_\l R_{i\a,j\b}=[R_{i\a,j\b},\lambda]. $$
Thus, we have
\eqn\swsolexpi{{\hat\o}_{i\a}=\o_{i\a}+{\sfrac{\hbar}{4}}C^{j\b,k\g} 
          \{\O_{j\b},\partial_{k\g}\o_{i\a}+R_{k\g,i\a}\}+\CO(\hbar^2).}

In order to find the field expansions for the remaining fields, we use the
constraints \constraintnc{} and the definitions \superspinor{}.
The expansion of \constraintnc{a} to first order in $\hbar$ leads directly to
$$\eqalignno{{\hat W}^1_{ij}&={\sfrac{1}{2}}\epsilon^{\a\b}\nabla_{(i\a}
            {\hat\o}^1_{j\b)}+{\sfrac{1}{8}}\epsilon^{\a\b}C^{m\d,n\epsilon}
            \{\partial_{m\d}\o_{i\a},\partial_{n\epsilon}\o_{j\b}\}, 
             &\swsol c\cr
}$$
where the parentheses mean normalized symmetrization. This solution can be
substituted into \constraintnc{b} to give the first order contribution of the
gauge potential ${\hat\bo}\ \!\!^i_\ad$. Assuming that 
$$ \bnabla^i_\ad {\hat\bo}\ \!\!^{j\, 1}_\bd=
   \bnabla^j_\bd {\hat\bo}\ \!\!^{i\, 1}_\ad, $$
we arrive at the equation
\eqn\efbo{\eqalign{
          \bnabla^i_\ad {\hat\bo}\ \!\!^{j\, 1}_\bd \ &=\ 
          {\sfrac{1}{2}}\epsilon_{\ad\bd}\epsilon^{ijkl}{\hat W}^1_{kl}
          + {\sfrac{1}{4}}C^{m\d,n\epsilon}\{\partial_{m\d}
            \bo^i_\ad,\partial_{n\epsilon}\bo^j_\bd\}\cr
          &=\ \Db^i_\ad{\hat\bo}\ \!\!^{j\, 1}_\bd -\{\bo^i_\ad,
             {\hat\bo}\ \!\!^{j\, 1}_\bd \}.\cr
}}
Remember that in the undeformed setup one considers
a certain gauge\foot{See also appendix A.}, namely
$\t\o-\bt\bo=\t^{i\a}\o_{i\a}+\bt^\ad_i\bo^i_\ad=0$. In the deformed regime we 
want to choose a similar gauge, 
\eqn\ogaugenc{\t{\hat\o}-\bt{\hat\bo}\ =\ 
              \t^{i\a}{\hat\o}_{i\a}+\bt^\ad_i{\hat\bo}\ \!\!^i_\ad\ =\ 0,}
implying that we have to all powers in $\hbar$
\eqn\ogaugenc{\t^{i\a}{\hat\o}_{i\a}^{(n)}+
              \bt^\ad_i{\hat\bo}\ \!\!^{i\, (n)}_\ad\ =\ 0.}
Considering this equation for $n=1$, one finds that the gauge potential
has to satisfy the relation
\eqn\conongp{{\hat\bo}\ \!\!^{i\, 1}_\ad=\Db^i_\ad(\t{\hat\o}^1)-
             \bt^\bd_j\Db^i_\ad{\hat\bo}\ \!\!^{j\, 1}_\bd.}
Substituting this equation into \efbo, we discover 
\eqn\efboi{\Db^i_\ad{\hat\bo}\ \!\!^{j\, 1}_\bd-\bt^\gd_l[\bo^i_\ad,
           \Db^j_\bd{\hat\bo}\ \!\!^{l\, 1}_\gd]=K^{ij}_{\ad\bd},}
where we have abbreviated
\eqn\efboii{K^{ij}_{\ad\bd}\equiv
            {\sfrac{1}{2}}\epsilon_{\ad\bd}\epsilon^{ijkl}{\hat W}^1_{kl}
          + {\sfrac{1}{4}}C^{m\d,n\epsilon}\{\partial_{m\d}\bo^i_\ad,
            \partial_{n\epsilon}
            \bo^j_\bd\}+\{\bo^i_\ad,\Db^j_\bd(\t{\hat\o}^1)\}.}
In order to simplify notation, let us introduce the shorthand index notation
${\bar A}=(\sm{\ \ \ad }{i })$, etc., and rewrite \efboi\ as
\eqn\efboiii{\Db_\Ab{\hat\bo}_\Bb^1-\bt^\Cb[\bo_\Ab,\Db_\Bb{\hat\bo}_\Cb^1]
             \ =\ K_{\Ab\Bb}.}
This equation might be iterated to give the solution for 
$\Db_\Ab{\hat\bo}_\Bb^1$,
\eqn\efboiv{\Db_\Ab{\hat\bo}_\Bb^1=\sum_{|{\bar I}|\leq 8}
             (-)^{\lfloor{|{\bar I}|\over 2}\rfloor}\,\bt^{{\bar I}}\,
             [\bo,K\}_{{\bar I},\Ab\Bb},}
where ``$\lfloor\ \rfloor$'' denotes the Gau{\ss} bracket and
\eqn\efbov{[\bo,K\}_{{\bar I},\Ab\Bb}\equiv[\bo_\Ab,[\bo_\Bb,[\bo_{\Ab_1},
           \cdots[\bo_{\Ab_{|{\bar I}|-2}},K_{\Ab_{|{\bar I}|-1}
           \Ab_{|{\bar I}|}}\}\cdots\}\}\}.}
Note that the sum in \efboiv\ is finite which is due to the nilpotency of
the $\bt$ in front of the bracket in \efboiii. Therefore the first order
contribution of the gauge potential ${\hat\bo}\ \!\!^i_\ad$ is given by
$$\eqalignno{{\hat\bo}_\Ab^1\ &=\ \Db_\Ab(\t{\hat\o}^1)-\bt^\Bb
               \sum_{|{\bar I}|\leq 8}
             (-)^{\lfloor{|{\bar I}|\over 2}\rfloor}\,\bt^{{\bar I}}\,
             [\bo,K\}_{{\bar I},\Ab\Bb}. &\swsol d\cr
}$$

Now we are able to write down the field expansions for the remaining fields.
The gauge potential ${\hat A}_{\a\bd}^1$ reads as
$$\eqalignno{{\hat A}_{\a\bd}^1\ &=\ {\sfrac{1}{8}}(\nabla_{i\a}
             {\hat\bo}\ \!\!^{i\, 1}_\bd-\bnabla^i_\bd{\hat\o}^1_{i\a}+
             {\sfrac{1}{2}}C^{m\d,n\epsilon}\{\partial_{m\d}\o_{i\a},
             \partial_{n\epsilon}\bo^i_\bd\}), &\swsol e\cr
}$$
which follows directly from \constraintnc{c}. The definitions \superspinor{}
finally give us
$$\eqalignno{{\hat\bchi}^1_{i\bd}\ &=\ -{\sfrac{1}{2}}\epsilon^{\a\b}(
             \nabla_{i\a}{\hat A}^1_{\b\bd}-\nabla_{\b\bd}{\hat\o}^1_{i\a}+
             {\sfrac{1}{2}}C^{m\d,n\epsilon}\{\partial_{m\d}\o_{i\a},
             \partial_{n\epsilon}A_{\b\bd}\}) , &\swsol f\cr
             {\hat\chi}\ \!\!^{i\, 1}_\b\ &=\ -{\sfrac{1}{2}}\epsilon^{\ad\bd}
             (\bnabla^i_\ad{\hat A}^1_{\b\bd}+\nabla_{\b\bd}
             {\hat\bo}\ \!\!^{i\, 1}_\ad-{\sfrac{1}{2}}C^{m\d,n\epsilon}
             \{\partial_{m\d}\bo^i_\ad,\partial_{n\epsilon}A_{\b\bd}\}). 
             &\swsol g\cr
}$$
Thus, we have computed the contributions to first order in the deformation.
Now one could proceed further and compute the higher order contributions.

{}Finally, the field expansions of the undeformed superfields, which are given 
in the appendix A, have to be substituted into \swsol{}. But this is not too
illuminating and we therefore refrain from doing this. We rather
concentrate ourselves on the zeroth order components, i.e., those which
do not contain the odd coordinates, since these will eventually give us the 
deformed field equations.

\subsec{Field equations}

The next thing we need to compute is the zeroth order components of the 
superfield equations \diracnc{} and \curveomnc{}. However, before we can 
derive them we have to discuss some preliminaries. In the equations \diracnc{} 
and \curveomnc{} products of the form $\t^I\star\t^J$ do appear.
Therefore we need to know their explicit zeroth order form. The 
first step in this direction is to show that
\eqn\wickthm{\eqalign{
             \t^{A_1}\star\cdots\star
            \t^{A_n}\ & =\ \t^{A_1}\cdots\t^{A_n}\ +\
             \sum\ {\rm all\ possible\ contractions} \cr
             & =\ \t^{A_1}\cdots\t^{A_n}\ +\ \sum_{i<j}\ 
             \t^{A_1}\cdots\contra{29}{\t^{A_i}\cdots\t}^{\!A_j}\cdots
             \t^{A_n}\ +\ \cdots,
}} 
with the indices $A_k=(i_k\a_k)$ and
\eqn\wickthmI{\contra{15}{\t^{A_i}\t}^{\!A_j}
               \equiv{\sfrac{\hbar}{2}}C^{A_iA_j}.}
Equation \wickthm\ resembles the fermionic Wick theorem. The signs have
to be taken as in the fermionic Wick theorem, i.e.,
$$\contra{29}{\t^{A_i}\t^{A_j}\t}^{\!A_k}\ =\ 
               -{\sfrac{\hbar}{2}}C^{A_iA_k}\t^{A_j},$$
for instance. The proof of \wickthm\ can easily be done by induction. 
For $n=2$, equation \wickthm\ is obviously satisfied. For $n>2$, one first 
shows that
\eqn\wickthmIII{(\t^{A_1}\cdots\t^{A_n})\star\t^{A_{n+1}}\ =\
                \t^{A_1}\cdots\t^{A_n}\t^{A_{n+1}}\ +\
                \sum_{i=1}^n\ \t^{A_1}\cdots\contra{33}{\t^{A_i}\cdots\
                \t}^{\!A_{n+1}},} 
from which then the assertion is immediate. 

Remember that any $\hf\in\IR^{(4|16)}_\hbar$ can be expanded in terms
of the odd coordinates. In order to simplify notation, let us define a 
projector $\pi_\circ$ projecting onto the zeroth order component, i.e., 
\eqn\zeroop{\pi_\circ\,:\,\hf(y,\t,\bt)\mapsto \nothi{\hf}(y)\ \ .}
Then we have
\eqn\zerocomp{\eqalign{
             \pi_\circ(\t^I\star\t^J)\ &=\ 
             \pi_\circ((\t^{A_1}\cdots\t^{A_n})\star 
             (\t^{B_1}\cdots\t^{B_m}))\cr
             &=\ \delta_{nm}{(-)^{{n\over 2}(n-1)
               } \hbar^n\over 2^n\, n!}\sum_{\{i,j\}}
             \epsilon_{i_1\cdots i_n}\epsilon_{j_1\cdots j_n}
             C^{A_{i_1}B_{j_1}}\cdots C^{A_{i_n}B_{j_n}}.
}}
The proof of \zerocomp\ is rather obvious. Starting point is the equation
\wickthmIII. It follows from this equation that at zeroth order only such 
terms of the product $\t^I\star\t^J$ contribute for which
$|I|=|J|$, because if $|I|\neq |J|$ equation \wickthmIII\ shows that there 
will be no fully contracted term and hence no contribution at zeroth order.
Then symmetry arguments, \wickthm\ and \wickthmIII\ lead to
$$ \pi_\circ((\t^{A_1}\cdots\t^{A_n})\star(\t^{B_1}\cdots\t^{B_n}))  
   = \left. \pi_\circ(\t^{A_1}\star\cdots\star\t^{A_n}\star\t^{B_1}\star
   \cdots\star\t^{B_n})
   \right|_{{\rm no\ }\contra{9}{A_iA}_{\!j},\ \contra{9}{B_iB}_{\!j}}, $$
i.e., the zeroth order component consists of all possible contractions but 
without contractions among the $\t^{A_i}$s ($\t^{B_i}$s).
Since from the very beginning the $\t^{A_i}$s ($\t^{B_i}$s) appear totally 
antisymmetrized, we obtain the Levi-Civita symbols on the right hand side of
\zerocomp. The factor of $1/n!$ provides the correct number of terms appearing 
after summation and the $2^n$ comes from the $n$ contractions. The sign 
$(-)^{{n\over 2}(n-1)}$ needs to be included, since in order to get the term 
proportional to $\epsilon_{i_1\cdots i_n}\epsilon_{j_1\cdots j_n}=
\epsilon_{1\cdots n}\epsilon_{1\cdots n}=1$ one has to anticommute 
${n\over 2}(n-1)$ thetas. As a corallary of \zerocomp\ it follows that
\eqn\zerocompI{\pi_\circ(\t^I\star\t^J)\ =\ \pi_\circ(\t^J\star\t^I).}

Let $\hf$ and $\hg$ be ${\frak g}$-valued elements of $\IR^{(4|16)}_\hbar$. 
In the equations \diracnc{} and \curveomnc{} always commutators or 
anticommutators of superfields appear. Therefore we are interested in 
$\pi_\circ([\hf,\hg\}_\star)$.
To compute this expression, we expand $\hf$ and $\hg$ as
\eqn\expanfg{\hf=\nothi{\ \hf}(y)\ \ + \sum_I {\hat f}_I(y)\t^I\ +\ \cdots
             \quad{\rm and}\quad
             \hg=\nothii{\hg}(y)\ \ + \sum_J {\hat g}_J(y)\t^J\ +\ \cdots,}
where the dots represent terms containing at least one $\bt$. Then we have to
distinguish three cases, namely $(p_\hf,p_\hg)=(0,0)$,  $(p_\hf,p_\hg)=(1,1)$ 
and $(p_\hf,p_\hg)=(0,1)$ leading to
\eqna\supercomexp
$$\eqalignno{\pi_\circ([\hf,\hg]_\star)\ = & 
             \ [\nothi{\hf},\nothii{\hg}]+\sum_{|I|=|J|}(-)^{p_I}
             [\hf_I,\hg_J]\ \pi_\circ(\t^I\star\t^J),
                                 & \supercomexp a\cr
            \pi_\circ(\{\hf,\hg\}_\star)\ = & \ 
              \{\nothi{\hf},\nothii{\hg}\}+\sum_{|I|=|J|}
             \{\hf_I,\hg_J\}\ \pi_\circ(\t^I\star\t^J),
                                 & \supercomexp b\cr
             \pi_\circ([\hf,\hg]_\star)\ = & 
             \ [\nothi{\hf},\nothii{\hg}]+\sum_{|I|=|J|}
             (\hf_I\hg_J-(-)^{p_I}\hg_J\hf_I)\ \pi_\circ(\t^I\star\t^J),
                                 & \supercomexp c\cr
}$$
respectively. In deriving these expressions we have used \zerocompI.

We have now all ingredients to give the zeroth order components of  
\diracnc{} and \curveomnc{}. For brevity, let us define 
$T^{IJ}\equiv\pi_\circ(\t^I\star\t^J)$. Remember that $T^{IJ}$ is
symmetric, i.e., $T^{IJ}=T^{JI}$. Putting everything together, the 
equations \diracnc{} become
\eqna\diracnczo
$$\eqalignno{\epsilon^{\a\b}\nhc_{\,\a\ad}\chc\ \!^i_\b +{\sfrac{1}{2}}
             \epsilon^{ijkl}[\Whc_{kl},\bchc_{j\ad}]  &= \cr 
             & \kern-3cm-\epsilon^{\a\b}\sum_{|I|=|J|}
               ({\hat A}_{\a\ad|I}{\hat\chi}\ \!\!_{\b|J}^i-(-)^{p_I}
               {\hat\chi}\ \!\!^i_{\b|J}{\hat A}_{\a\ad|I})\,T^{IJ}
             & \diracnczo a\cr                   
             -{\sfrac{1}{2}}\epsilon^{ijkl}
             &\sum_{|I|=|J|} 
               ({\hat W}_{kl|I}{\hat\bchi}_{j\ad|J}-(-)^{p_I}
               {\hat\bchi}_{j\ad|J}{\hat W}_{kl|I})\,T^{IJ}, \cr
             \epsilon^{\ad\bd}\nhc_{\,\a\ad}\bchc_{i\bd}+[\Whc_{ij},
             \chc\ \!^j_\a]
             & = \cr
             & \kern-3cm-\epsilon^{\ad\bd}\sum_{|I|=|J|}
               ({\hat A}_{\a\ad|I}{\hat\bchi}_{i\bd|J}-(-)^{p_I}
               {\hat\bchi}_{i\bd|J}{\hat A}_{\a\ad|I})\,T^{IJ}
             & \diracnczo b\cr                   
           - & \sum_{|I|=|J|}
               ({\hat W}_{ij|I}{\hat\chi}\ \!\!^j_{\a|J}-(-)^{p_I}{\hat\chi}\ 
               \!\!^j_{\a|J}{\hat W}_{ij|I})\,T^{IJ}, \cr
}$$
while the equations of motion for the gauge field and the scalar multiplet are
\eqna\curveomnczo
$$\eqalignno{\epsilon^{\ad\bd}\nhc_{\g\ad}\nothi{\hf}_{\bd\gd}+\epsilon^{\a\b}
               \nhc_{\a\gd}\nothi{\hf}_{\b\g}-{\sfrac{1}{4}}\epsilon^{ijkl}
             &
               [\nhc_{\g\gd}\Whc_{ij},\Whc_{kl}] -\{\chc\ \!^i_\g,
               \bchc_{i\gd}\}= 
             & \cr                           
             & \kern-4cm-\sum_{|I|=|J|}(-)^{p_I}\left\{\epsilon^{\ad\bd}
               [{\hat A}_{\g\ad|I},\hf_{\bd\gd|J}]+\epsilon^{\a\b}
               [{\hat A}_{\a\gd|I},\hf_{\b\g|J}]\right\}\,T^{IJ} 
             & \curveomnczo a\cr
             & \kern-3cm+\sum_{|I|=|J|}\left\{(-)^{p_I}{\sfrac{1}{4}}
               \epsilon^{ijkl}[(\hnab_{\g\gd}{\hat W}_{ij})_I,{\hat W}_{kl|J}]+
               \{{\hat\chi}\ \!\!^i_{\g|I},{\hat\bchi}_{i\gd|J}\}\right\}\,
               T^{IJ},
             & \cr
               \epsilon^{\a\b}\epsilon^{\ad\bd}\nhc_{\a\ad}
               \nhc_{\b\bd}\Whc_{ij}
               -{\sfrac{1}{4}}\epsilon^{klmn}[\Whc_{mn},&[\Whc_{kl},
               \Whc_{ij}]]  
               ={\sfrac{1}{2}}\epsilon_{ijkl}\epsilon^{\a\b}
               \{\chc\ \!^k_\a,\chc\ \!^l_\b\}+\epsilon^{\ad\bd}\{\bchc_{i\ad},
               \bchc_{j\bd}\}
             & \cr
             & \kern-5.2cm-\sum_{|I|=|J|}(-)^{p_I}\left\{\epsilon^{\a\b}
               \epsilon^{\ad\bd}[{\hat A}_{\a\ad|I},(\hnab_{\b\bd}
               {\hat W}_{ij})_J]
               -{\sfrac{1}{4}}\epsilon^{klmn}[{\hat W}_{mn|I},
               [{\hat W}_{kl},{\hat W}_{ij}]_J]
               \right\}\,T^{IJ}
             & \curveomnczo b\cr
             & \kern-2.7cm+\sum_{|I|=|J|}\left\{{\sfrac{1}{2}}\epsilon_{ijkl}
               \epsilon^{\a\b}\{{\hat\chi}\ \!\!^k_{\a|I},{\hat\chi}\ 
               \!\!^l_{\b|J}\}+\epsilon^{\ad\bd}
               \{{\hat\bchi}_{i\ad|I},{\hat\bchi}_{j\bd|J}\}\right\}\,T^{IJ}.
             & \cr
}$$
These equations and the field expansions \swsol{} together with the undeformed
superfield expansions given in the appendix A allow us to write down the
deformed field equations. The derivations of the zeroth order components
of \swsol{} is pretty lengthy but straightforward. We therefore will not
present them here, but only give the results. We ultimately find
\eqna\swsolzo
$$\eqalignno{\Whc_{ij}\ &=\ \Wc_{ij}+{\sfrac{\hbar}{2}}\,C^{m\d,n\epsilon}\,
             \epsilon_{\d\epsilon}\{\Wc_{mi},\Wc_{jn}\}+ \CO(\hbar^2), 
              & \swsolzo a\cr
             \nothi{\hat{A}}_{\a\bd}\ &=\ 
             \nothiii{A}_{\a\bd}+{\sfrac{\hbar}{4}}\,C^{m\d,n\epsilon}\,
             \epsilon_{\a\d}\{\Wc_{mn},\nothiii{A}_{\epsilon\bd}\}
             +\CO(\hbar^2), & \swsolzo b\cr
             \bchc_{i\bd}\ &=\ \bcc_{i\bd}+{\sfrac{\hbar}{96}}\,
             C^{m\d,n\epsilon}\,[
             11\epsilon_{\d\epsilon}(\{\Wc_{mn},\bcc_{i\bd}\}-
             2\{\Wc_{in},\bcc_{m\bd}\})
             &{}\cr
             &\kern1.5cm
             -5(\epsilon_{mnij}\{\nothiii{A}_{\d\bd},\cc\ \!\!^j_\epsilon\})
             ]+\CO(\hbar^2),& \swsolzo c\cr
             \chc\ \!\!^i_\b\ &=\  \cc\ \!\!^i_\b+{\sfrac{\hbar}{16}}\, 
             C^{m\d,n\epsilon}\,[\{\Wc_{mn},{\sfrac{4}{3}}\epsilon_{\epsilon\d}
             \cc\ \!\!^i_\b-{\sfrac{11}{3}}\epsilon_{\d\b}
             \cc\ \!\!^i_\epsilon\}&{}\cr
             &\kern1.5cm-\d^i_m\{\Wc_{ln},{\sfrac{4}{3}}\epsilon_{\epsilon\d}
             \cc\ \!\!^l_\b+{\sfrac{7}{3}}\epsilon_{\epsilon\b}
             \cc\ \!\!^l_\d-{\sfrac{2}{3}}\epsilon_{\d\b}\cc\ \!\!^l_\epsilon\}
             &{}\cr
             &\kern1.5cm-\epsilon_{\b\epsilon}\epsilon^{\ad\bd}
             \{\nothiii{A}_{\d\ad},12\d^i_m\bcc_{n\bd}-
             {\sfrac{1}{2}}\d^i_n\bcc_{m\bd}\}]+\CO(\hbar^2).
             & \swsolzo d\cr
}$$

Now one substitutes these expressions into the equations \diracnczo{} and
\curveomnczo{} and uses the undeformed expansions given in the appendix A
to obtain the deformed super Yang-Mills equations to first order
in $\hbar$, e.g.,
$\epsilon^{\a\b}\nc_{\a\ad}\cc\ \!\!^i_\b+{\sfrac{1}{2}}
\epsilon^{ijkl}[\Wc_{kl},\bcc_{j\ad}]  = \CO(\hbar)$
(note that solving this equation consistently together
with the other equations of 
motion makes obviously the fields on the left-hand side $\hbar$-dependent).
Actually performing this task leads to both unenlightening and
complicated looking expressions, so we refrain from writing them down.
To proceed in a realistic manner, one can constrain the deformation parameters
to obtain manageable equations of motion.

For instance, in order to compare the deformed equations of motion with 
Seiberg's deformed $\CN=1$ equations\foot{or similarly in the case of the
deformed $\CN=2$ equations in $\CN=1$ superspace language \ArakiSE}
\SeibergYZ, one would have to 
restrict the deformation matrix $C^{i\a,j\b}$ properly and to put 
some of the fields, e.g., $W_{ij}$, to zero. Additionally, one would 
have to rotate the 
fermion field content such that the symplectic reality condition induced by
\rc{} holds. Here, however, one encounters
the subtlety that on $\CN=1$ superspace with Euclidean signature 
the gauge potentials are necessarily complex (cf., e.g., reference 
\LukierskiJW). For these and other reasons, this comparison would
carry us too far afield from the main thread of development of the present
paper. Therefore, we shall discuss this issue in our forthcoming work \Saemann.

\newsec{Conclusions}

In this paper we have proposed a way of deforming $\CN=4$
super Yang-Mills theory. The starting point was the constraint 
equations on the deformed superspace $\IR_\hbar^{(4|16)}$ from which we
derived the deformed superspace equations of motion. By using a
generalization of the Seiberg-Witten map to superspace, 
we gave a systematic procedure of 
constructing the deformed superfields order by order in the deformation 
$\hbar$. Eventually, these yield deformed equations of motion on $\IR^4$ 
with a larger number of deformation parameters than in the case of $\CN=1,2$
deformations.

Generalizing the string theory side of the derivation of Seiberg-Witten maps 
seems to be nontrivial. The graviphoton used to deform the fermionic
coordinates belongs to the R-R sector, while the gauge field strength
causing the deformation in the bosonic case sits in the 
NS-NS sector. This implies, that the field strengths appear on different
footing in the vertex operators of the appropriate string theory (type II
with $\CN=2$, $d=4$). A first step might be to consider a ``pure gauge'' 
configuration in which the gluino and gluon field strengths vanish. The 
corresponding vertex 
operator in Berkovits' hybrid formalism on the boundary of the worldsheet of
an open string contains the terms \BerkovitsUE,
$$V={\sfrac{1}{2\a'}}\int d\tau\ (\dot{\t}^\a\o_\a+\dot{X}^\mu A_\mu
             -i\sigma^\mu_{\a\ad}\dot{\t}^\a\bt^\ad A_\mu),$$
with the formal (classical) gauge transformations $\d_\l\o_\a=D_\a\l$ and 
$\d_\l A_\mu=\partial_\mu\l$. From here, one may proceed exactly as in 
\SeibergVS\ using the deformation of \SeibergYZ: regularization of the 
action by Pauli-Villars\foot{Pauli-Villars
was applied to supergravity, e.g., in \GaillardMN.} and point-splitting 
procedures lead
to an undeformed and a deformed gauge invariance, respectively. Although on
flat Euclidean space, pure gauge is trivial, the two different gauge 
transformations obtained are not.

More general, a Seiberg-Witten map is a translation rule between two physically
equivalent field theories. The fact that our choice of the deformation
\defp\ generically breaks half of the supersymmetry is not in contradiction
with the existence of a Seiberg-Witten map, but may be seen analogously to 
the purely bosonic case: in both the commutative and noncommutative theories,
particle Lorentz invariance is broken which is due to the background field 
($B$-field).

Furthermore, there are several open questions which should be clarified. 
We only indicated the construction of the deformed superfields, since already
to first order in $\hbar$ the computations became pretty lengthy. 
Therefore the question is whether the changes in the deformed superfields
in comparison to the undeformed ones are polynomial in $\hbar$ as 
suggested by the nilpotency of the star product. On the other hand,
it remains to clarify whether the one-to-one correspondence between
the deformed equations of motion and the constraint equations is still
valid on $\IR^{(4|16)}_\hbar$. 

Another point concerns the Seiberg-Witten map proposed above: it might be used
to shed light on the question of supersymmetry breaking in different approaches
to the deformation of the fermionic coordinates. E.g., in \SeibergYZ, the
approach also used above, the deformation breaks supersymmetry in general, 
while \FerraraMM\ uses a supersymmetric deformation preserving the algebra at
the price of losing chirality. In the latter case, one should also be able 
to construct a Seiberg-Witten map. However, here the ${\bar D}$s are no longer
derivations with respect to the star product and hence they should be modified
properly (by using Kontsevich's formality map) in 
order to follow the same steps as presented above. For this 
approach, the Seiberg-Witten map would relate two (fully) supersymmetric
theories. A third point
of view is found in \OoguriQP: a deformation of the fermionic coordinates is 
also induced by a self-dual graviphoton background but later on compensated by
introducing a gluino background so that the ordinary superspace is restored.

Finally,
it would be illuminating to explore the connection of the constraint
equations and the auxiliary linear system of partial differential equations
(to which the constraint equations are the compatibility condition) on
the deformed superspace. In the undeformed case, the existence of such a 
linear 
system \refs{\VolovichKR,\AbdallaXQ,\ChauIV} leads to the 
integrability of $\CN=4$ super Yang-Mills theory. Therefore dressing
\Zakharov\ and splitting \WardTA\ methods can be applied 
\refs{\WittenNT,\Kapranov,\GervaisHH,\GervaisVJ} for its solving. 
It would be interesting to generalize these methods not only to the
noncommutative case as in 
\refs{\LechtenfeldAW,\LechtenfeldGF,\LechtenfeldIE,\WolfJW,\HorvathBJ,
\IhlKZ,\LechtenfeldVV}, but also to the nonanticommutative deformation
of $\CN=4$ super Yang-Mills theory.

\bigbreak\bigskip\bigskip\centerline{{\bf Acknowledgements}}\nobreak
We thank A. Kling, O. Lechtenfeld, A. D. Popov and S. Uhlmann for useful 
comments. We are also very grateful to B. Zupnik for pointing out
an essential mistake in the first version of this paper which eventually
led to major changes. This work was done within the framework
of the DFG priority program (SPP 1096) in string theory.

\appendix{A}{Superfield expansions for vanishing deformation}

In this appendix we shall review how one can construct the superfields 
from their leading components. Following \refs{\HarnadVK,\HarnadBC}, we 
impose the gauge condition
\eqn\gaugeA{\t\o-\bt\bo\ =\ \t^{i\a}\o_{i\a}+\bt^\ad_i\bo^i_\ad\ =\ 0}
in order to remove the superfluous gauge degrees of freedom associated with
the $\t^{i\a}$ and $\bt^\ad_i$ coordinates. Moreover, we need some 
recursion operator, $\CD$, which leads to the proper field expansions.
We take the following form \refs{\HarnadVK,\HarnadBC}
\eqn\recopA{\eqalign{
             \CD f\ &\equiv\ (\t D+\bt\Db)f \cr
             & =\ (\t^{i\a}D_{i\a}+\bt_{i\ad}\Db^{i\ad})f\
             =\ (\t^{i\a}D_{i\a}-\bt_i^\ad\Db^i_\ad)f,\cr
}}
where $f\in\IR^{(4|16)}$. 
It then follows immediately that in the gauge \gaugeA\ the recursion operator 
$\CD$ is the same as the covariant one, i.e.,
\eqn\recopAi{\CD\ =\ \t\nabla+\bt\bnabla.}

By using the undeformed version of the constraint equations \constraintnc{}, 
we obtain after some simple algebraic manipulations 
\eqna\recursionA
$$\eqalignno{(1+\CD)\o_{i\a}\ =  & \ 2\bt^\ad_i A_{\a\ad}-2\epsilon_{\a\b}
             \t^{j\b}W_{ij},    & \recursionA a\cr
             (1+\CD)\bo^i_\ad\ = & \ 2\t^{i\a}A_{\a\ad}-\epsilon_{\ad\bd}
             \epsilon^{ijkl}\bt^\bd_jW_{kl}. 
                                 & \recursionA b\cr
}$$
Finally, the (undeformed) equations
\superspinor{}, \supereqI{} and \supereqII{} give us the remaining
relations
$$\eqalignno{\CD A_{\a\ad}\ =   & \ -\epsilon_{\a\b}\t^{i\b}\bchi_{i\ad}+
             \epsilon_{\ad\bd}\bt^\bd_i\chi^i_\a,
                                & \recursionA c\cr
         \CD W_{ij}\ =      & \ \epsilon_{ijkl}\t^{k\a}\chi^l_\a
             -\bt^\ad_i\bchi_{j\ad}+\bt^\ad_j\bchi_{i\ad},
                                & \recursionA d\cr
             \CD\chi^i_\a\ =    & \ -2\t^{i\b}f_{\a\b}+{\sfrac{1}{2}}
             \epsilon_{\a\b}\epsilon^{iklm}\t^{j\b}[W_{lm},W_{jk}]-
             \epsilon^{ijkl}\bt^\ad_j\nabla_{\a\ad}W_{kl}, 
                                & \recursionA e\cr
             \CD\bchi_{i\ad}\ = & \ 2\t^{j\a}\nabla_{\a\ad}W_{ij}+
             2\bt^\bd_i f_{\ad\bd}+{\sfrac{1}{2}}\epsilon_{\ad\bd}
             \epsilon^{jklm}\bt^\bd_j[W_{lm},W_{ik}],
                                & \recursionA f\cr
}$$
as one may readily verify. The equations \recursionA{} are regarded as a
recursive definition of the superfields, i.e., by iterating these equations we
can reconstruct the superfields order by order in the odd coordinates from
their leading components. To exemplify our situation, let us write down the
expansions of the superfields $A_{\a\ad}$, $W_{ij}$, $\chi^i_\a$ and 
$\bchi_{i\ad}$ up to quadratic order in $\t$ (no $\bt$s),
\eqna\sfeA
$$\eqalignno{\kern-1.2cm A_{\a\ad} & = 
             \nothiii{A}_{\a\ad}+\epsilon_{\a\b}\bcc_{i\ad}\t^{i\b}
             -\epsilon_{\a\b}\nc_{\a\ad}\Wc_{ij}\t^{i\b}\t^{j\g}
             \ +\ \cdots, &  \sfeA a\cr   
             \kern-1.2cm W_{ij}    & = 
             \Wc_{ij}-\epsilon_{ijkl}\cc\ \!^l_\a\t^{k\a}-
             \epsilon_{ijkl}(\delta^l_m\nothiii{f}_{\!\b\a}+{\sfrac{1}{4}}
             \epsilon_{\b\a}\epsilon^{lnij}[\Wc_{ij},\Wc_{mn}])\t^{k\a}
             \t^{m\b}\ +\ \cdots,
                                 & \sfeA b\cr
             \kern-1.2cm\chi^i_\a & =
             \cc\ \!^i_\a-(2\delta^i_j\nothiii{f}_{\!\b\a}+{\sfrac{1}{2}}
             \epsilon_{\b\a}\epsilon^{iklm}[\Wc_{lm},\Wc_{jk}])\t^{j\b}\ + &\cr
             &\kern.6cm \left\{{\sfrac{1}{2}}
             \delta^i_j\epsilon^{\ad\bd}(\epsilon_{\g\a}
             \nc_{\b\ad}\bcc_{k\bd}+\epsilon_{\g\b}\nc_{\a\ad}\bcc_{k\bd})\ -
             \right.\cr
             &\kern1.6cm\left.
             {\sfrac{1}{4}}\epsilon_{\a\b}\epsilon^{ipmn}(\epsilon_{jkpq}
             [\Wc_{mn},\cc\ \!^q_\g]+\epsilon_{mnkq}[\Wc_{jp},\cc\ \!^q_\g])
             \right\}\t^{j\b}\t^{k\g}\ +\ \cdots,
                                   & \sfeA c\cr
             \kern-1.2cm\bchi_{i\ad} & =
             \bcc_{i\ad}+2\nc_{\a\ad}\Wc_{ij}\t^{j\a}+(\epsilon_{ijkl}
             \nc_{\a\ad}\cc\ \!^l_\b+\epsilon_{\a\b}[\Wc_{ij},\bcc_{k\ad}])
             \t^{j\a}\t^{k\b}\ +\ \cdots,
                                 & \sfeA d\cr
}$$
where 
$$ \nothiii{f}_{\!\a\b}=-{\sfrac{1}{2}}\epsilon^{\ad\bd}[\nc_{\a\ad},
   \nc_{\b\bd}]
   =-{\sfrac{1}{2}}\epsilon^{\ad\bd}(\partial_{\a\ad}\nothiii{A}_{\b\bd}-
    \partial_{\b\bd}\nothiii{A}_{\a\ad}+[\nothiii{A}_{\a\ad},
    \nothiii{A}_{\b\bd}]). 
$$ 
To arrive at \sfeA{} we have used the formal field expansion 
\expan\ and the equations \recursionA{c-f}. It is important to stress that
the recursions do not involve the field equations. Therefore all superfields
are well defined off-shell.

Now we are able to give the expansions of $\o_{i\a}$ and $\bo^i_\ad$. The
equations \recursionA{a,b} then yield
$$\eqalignno{\o_{i\a} \ &=\ -\epsilon_{\a\b}\Wc_{ij}\t^{j\b}+
                            \d^j_i\nothiii{A}_{\a\ad}\bt^\ad_j-{\sfrac{2}{3}}
                           \epsilon_{\a\b}\epsilon_{ijkl}\cc\ \!\!^l_\d\t^{j\b}
                         \t^{k\d} &{}\cr
                &\kern1cm+{\sfrac{2}{3}}\epsilon_{\a\g}(2\d^l_i
                          \bcc_{k\gd}-\d^l_k\bcc_{i\gd})\t^{k\g}\bt^\gd_l
                           +{\sfrac{2}{3}}\d^j_i\epsilon_{\ad\bd}
                           \cc\ \!\!^k_\a\bt^\ad_j\bt^\bd_k\ +\ 
                          \cdots,  & \sfeA e\cr
               \bo^i_\ad\ &=\ \d^i_j \nothiii{A}_{\a\ad}\t^{j\a}-{\sfrac{1}{2}}
               \epsilon_{\ad\bd}\epsilon^{ijkl}\Wc_{kl}\bt^\bd_j 
          -{\sfrac{2}{3}}\d^i_j\epsilon_{\a\b}\bcc_{k\ad}\t^{j\a}\t^{k\b}&{}\cr
        &\kern1cm+{\sfrac{2}{3}}\epsilon_{\ad\gd}(2\d^i_k\cc\ \!\!^l_\g-\d^l_k
              \cc\ \!\!^i_\g)\t^{k\g}\bt^\gd_l+{\sfrac{2}{3}}\epsilon_{\ad\bd}
              \epsilon^{ijkl}\bcc_{l\gd}\bt^\bd_j\bt^\gd_k\ +\ \cdots .
             & \sfeA f \cr
}$$
Here, we have also written down the $\bt$, $\t\bt$ and $\bt\bt$ components as 
we need them in the discussion of the deformed superfields.

 \listrefs
\end